\documentclass[11pt]{article}
\usepackage{graphicx, amssymb}
\newcommand{\SetFigFont}[3]{}

\title{A Time Independent Energy Estimate for Outgoing Scalar Waves
in the Kerr Geometry}
\author{Felix Finster\thanks{Research supported in part by the Deutsche
Forschungsgemeinschaft.} and Joel Smoller\thanks{Research supported in
part by the Humboldt Foundation and the National Science Foundation,
Grant No.~DMS-0603754.}}
\date{July 2007}

\newtheorem{Def}{Def.}[section]
\newtheorem{Thm}[Def]{Theorem}
\newtheorem{Prp}[Def]{Proposition}
\newtheorem{Lemma}[Def]{Lemma}

\newtheorem{Corollary}[Def]{Corollary}
\newcommand{\Proof}{{\em{Proof. }}}
\newcommand{\QED}{\ \hfill $\FBox$ \\[1em]}
\newcommand{\spc}{\;\;\;\;\;\;\;\;\;\;}
\newcommand{\bra}{\mbox{$< \!\!$ \nolinebreak}}
\newcommand{\ket}{\mbox{\nolinebreak $>$}}
\newcommand{\lbra}{\langle}
\newcommand{\lket}{\rangle}
\newcommand{\C}{\mathbb{C}}
\newcommand{\R}{\mathbb{R}}

\newcommand{\Z}{\mathbb{Z}}

\newcommand{\N}{\mathbb{N}}
\newcommand{\sN}{\mbox{\rm \scriptsize I \hspace{-.8 em} N}}

\newcommand{\beq}{\begin{equation}}
\newcommand{\eeq}{\end{equation}}

\newcommand{\FBox}{\rule{2mm}{2.25mm}}

\newcommand{\low}{{\mbox{\tiny{$<$}}}}
\newcommand{\high}{{\mbox{\tiny{$>$}}}}
\newcommand{\cin}{c_{\mbox{\scriptsize{\rm{in}}}}}
\newcommand{\cout}{c_{\mbox{\scriptsize{\rm{out}}}}}

\begin{document}
\maketitle

\begin{abstract}
The Cauchy problem for the scalar wave equation in the
Kerr geometry is considered, with initial data which is smooth and compactly supported
outside the event horizon.
A time-independent energy estimate for the outgoing wave is obtained.
As an application we estimate the outgoing energy for
wave-packet initial data, uniformly as the support of the initial data is shifted
to infinity. The main mathematical tool is our previously
derived integral representation of the wave propagator.
\end{abstract}

\section{Introduction}
\setcounter{equation}{0}
In the recent papers~\cite{FKSY, FKSY2, FKSY3} the Cauchy problem for the scalar wave equation
in the Kerr geometry was studied for smooth initial data with compact support outside the
event horizon. It was proved that the wave function decays in
$L^\infty_{\mbox{\scriptsize{loc}}}$. This implies that the wave splits into an ingoing
and outgoing wave, which enters the the black hole and escapes to infinity, respectively.
Due to energy conservation, the total energy of the ingoing and outgoing waves
is time-independent. However, since the energy density need not be positive inside
the ergosphere, the energy of the ingoing wave may be negative,
and thus energy conservation does not give an energy estimate for the outgoing wave.
In the present paper we show that despite this difficulty, the energy of the outgoing wave is
nevertheless bounded uniformly in time.

In the spherically symmetric Schwarzschild geometry, the physical energy density
of scalar waves is non-negative, and thus one can estimate the wave using
energy estimates~\cite{KW}. In the Kerr geometry, the rotation of the black hole
leads to the couter-intuitive effect that the energy density can be negative in a
region near and outside the event horizon called the ergosphere.
This effect has interesting consequences; in particular, it gives rise to a mechanism
of extracting energy from a rotating black hole. This was discovered by Roger
Penrose~\cite{Penrose}, who considered a classical point particle which enters
the ergosphere where it disintegrates into two partices.
The wave analogue of the Penrose process was proposed by Zeldovich~\cite{Z}
and studied numerically by Starobinski~\cite{S} on the level of modes.
In the recent paper~\cite{FKSY4}, we give a more convincing argument in terms of solutions of the Cauchy problem for wave packet initial data.  In order to make this argument completely rigorous, it is necessary to obtain time independent energy estimates for outgoing scalar waves in the Kerr geometry,  independent of the wave packet, as the wave packet initial data is moved to infinity.
Due to the negative energy density inside the ergosphere, energy estimates do not apply.
The results of~\cite{FKSY, FKSY2, FKSY3}, where an integral representation for
solutions of the scalar wave equation is derived and used to prove local decay,
are not good enough, because in these papers the wave is estimated only in a compact
region of space, but not in the unbounded region near infinity. In the present paper we
develop a technique for estimating the scalar wave globally in space, uniformly in time.

A rotating black hole is modeled by the Kerr metric, which in
Boyer-Lindquist coordinates takes the form
\begin{eqnarray}
\lefteqn{ ds^2 \;=\; g_{jk}\:dx^j\, dx^k } \nonumber \\
&=& \frac{\Delta}{U} \:(dt \:-\: a \:\sin^2 \vartheta \:d\varphi)^2
\:-\: U \left( \frac{dr^2}{\Delta} + d\vartheta^2 \right) \:-\:
\frac{\sin^2 \vartheta}{U} \:(a \:dt \:-\: (r^2+a^2) \:d\varphi)^2 .\quad
\label{eq:0}
\end{eqnarray}
Here $M>0$ and $aM>0$ denote the mass and the angular momentum of the black hole,
respectively, and
the functions~$U$ and $\Delta$ are given by
\[ U(r, \vartheta) \;=\; r^2 + a^2 \:\cos^2 \vartheta \;,\spc
\Delta(r) \;=\; r^2 - 2 M r + a^2  \: . \]
We consider only the {\em{non-extreme case}} $M^2 > a^2$, where the function $\Delta$
has two distinct roots. The largest root
\[ r_1 \;=\; M \:+\: \sqrt{M^2 - a^2} \]
defines the event horizon of the black hole.
We restrict attention to the region $r>r_1$ outside the event horizon.
The scalar wave equation is
\begin{equation} \label{swave}
\square \,\Phi:= g^{ij}\nabla_{i}\nabla_{j}\,\Phi \;=\; 0 \:,
\end{equation}
where~$\nabla$ denotes covariant differentiation. We fix initial data~$\Psi_0 \in C^\infty_0((r_1, \infty)
\times S^2)^2$ and decompose the $\varphi$-dependence in a Fourier series,
\beq \label{fs}
\Psi_0(r,\vartheta, \varphi) \;=\;
\sum_{k \in \Z} \Psi_0^k(r, \vartheta)\: e^{-i k \varphi} \:.
\eeq
Since the $k$-modes propagate independently, we can restrict
attention to a fixed~$k$-mode. The main result of this paper
is the following theorem.
\begin{Thm} \label{thm11}
Consider the Cauchy problem
\begin{equation} \label{cauchy}
\square\,\Phi \;=\;0 \:,\spc
(\Phi, i \partial_{t}\Phi)_{|t=0} \;=\; \Psi^k_0(r, \vartheta) \: e^{-i k \varphi}
\end{equation}
for initial data $\Psi^k_0 \in C^\infty_0((r_1, \infty) \times S^2)^2$,
whose $\varphi$-dependence is given by $e^{-i k \varphi}$.
Then for any~$R>r_1$ there is a constant~$C=C(R, \Psi_0)$ such that
the energy of~$\Psi(t)$ outside a neighborhood of the event horizon
is bounded uniformly in time,
\beq \label{energybound}
\int_{[R, \infty) \times S^2} {\mathcal{E}}(\Phi(t,x))\, dr\, d\varphi\, d\cos \vartheta
\;\leq\; C \:.
\eeq
Here~${\mathcal{E}}$ denotes the energy density (see~(\ref{ED})).
\end{Thm}
Note that we make no statement on the precise dependence of the constant~$C$ on the initial data.
The energy bound~(\ref{energybound}) implies that the $L^2$-norm of the first
derivatives of~$\Phi$ on~$\R^3 \setminus B_R(0)$
are bounded (with respect the usual integration
measure $r^2 dr \, d\varphi\, d\cos \vartheta$). In the case~$k \neq 0$,
we obtain furthermore a weighted~$L^2$-estimate of the form
\[ \int_{\R^3 \setminus B_R(0)} |\Phi(t,x)|^2\: \frac{d^3x}{|x|^2}
\;\leq\; C\:. \]
Since the time derivatives of~$\Phi$ also satisfy the scalar wave equation,
our theorem yields similar estimates for all time derivatives of~$\Phi$.
Furthermore, using that the spatial part of the wave equation~(\ref{swave}) is elliptic,
we immediately obtain similar bounds for all spatial derivatives.
Hence in the case~$k \neq 0$,
our theorem gives time-independent bounds for all corresponding weighted Sobolev norms of~$\Phi$.

In the rigorous treatment of superradiance in~\cite{FKSY4} a slightly different version of Theorem~\ref{thm11}
is needed. Namely, for wave-packet initial data one needs to control the energy near infinity for large time, 
uniformly as the the initial data is shifted towards infinity.
The last section in this paper is devoted to such an energy estimate, see Theorem~\ref{thm2}.

\section{Preliminaries}
\setcounter{equation}{0}
In this section we recall basic facts and collect useful formulas;
the details can be found in~\cite{FKSY, FKSY2}.
We consider for given~$\Psi_0 \in C^\infty_0((r_1, \infty) \times S^2)$
a summand~$\Psi_0^k$ in the Fourier expansion~(\ref{fs}).
In the case~$k=0$, the energy density is positive, and thus
Theorem~\ref{thm11} follows immediately from energy conservation. Possibly after a time reversal
and the replacement~$k \rightarrow -k$, we may
assume that~$k>0$. For simplicity in notation, we usually omit
the index~$k$ and the~$\varphi$-dependence.

Setting~$\Psi=(\Phi, i \partial_t \Phi)$, the wave equation can be written in
{\em{Hamiltonian form}}
\[ i\partial_{t}\Psi \;=\; H\,\Psi\:, \]
where the operator~$H$ is given by
\beq \label{Hdef}
H \;=\; \left(
\begin{array}{cc}
0 & 1 \\ A & \beta \end{array} \right) ,
\eeq
with~$A$ an elliptic operator and~$\beta$ a multiplication operator,
\begin{eqnarray}
A &=& \frac{1}{\rho} \left[-\frac{\partial}{\partial u} (r^2+a^2)
\frac{\partial}{\partial u} - \frac{\Delta}{r^2+a^2}\: \Delta_{S^2}
- \frac{a^2 k^2}{r^2+a^2} \right] \label{Adef} \\
\beta &=& -\frac{2 a k}{\rho} \left( 1 - \frac{\Delta}{r^2+a^2} \right) \\
\rho &=& r^2+a^2 \:-\: a^2\: \sin^2 \vartheta\: \frac{\Delta}{r^2+a^2} \:,
\end{eqnarray}
and~$\Delta_{S^2}$ denotes the Laplacian on~$S^2$.
Here~$u \in \R$ is the {\em{Regge-Wheeler coordinate}} defined by
\begin{equation} \label{51a}
\frac{du}{dr} \;=\; \frac{r^2+a^2}{\Delta} \;,\spc
\frac{\partial}{\partial r} \;=\; \frac{r^2+a^2}{\Delta}\: \frac{\partial}{\partial u} \:,
\end{equation}
mapping the event horizon~$r=r_1$ to~$u=-\infty$.
The physical {\em{energy density}} is given by~\cite[eq.\ (2.13)]{FKSY}
\begin{eqnarray}
{\mathcal{E}} &=&
\left({\frac{(r^{2}+a^{2})^{2}}{\Delta}} - a^{2}\,\sin^{2}\vartheta \right) |\partial_{t} \Phi|^2
\:+\: \Delta \, |\partial_{r} \Phi|^2 \nonumber \\
&& +\sin^{2}\vartheta\, |\partial_{\cos \vartheta}\Phi|^2
+\left( {\frac{k^2}{\sin^{2}\vartheta}}-{\frac{a^{2} k^2}{\Delta}}\right)\,|\Phi|^2 \:.
\label{ED}
\end{eqnarray}
This energy density need not be positive everywhere. Indeed, it can be
negative inside the {\em{ergosphere}}, defined by the
inequality~$\Delta - a^2 \sin^2 \vartheta < 0$.

Integrating the corresponding bilinear form over space gives the {\em{energy inner product}},
which in the Regge-Wheeler coordinate can be written as follows,
\beq \label{eipr}
\bra \Psi_1, \Psi_2 \ket \;:=\; \int_{\R \times S^2}
\lbra \Psi_1, \left( \begin{array}{cc} A & 0 \\ 0 & 1 \end{array} \right)
\Psi_2 \lket_{\C^2} \:\rho\: du\: d(\cos \vartheta)\:.
\eeq
The Hamiltonian~$H$ is symmetric with respect to this inner product.

As in~\cite[Section~1]{FKSY} the wave equation can be separated with the
multiplicative ansatz
\beq \label{Phisep}
\Phi(t,r,\vartheta) \;=\; e^{-i \omega t}\: R(r)\, \Theta(\vartheta)
\eeq
with a real parameter~$\omega$, giving rise to the system of ODEs
\beq \label{sepwave}
{\cal{R}}_\omega\,R_\lambda \;=\; -\lambda \, R_{\lambda},\qquad  {\cal{A}}_\omega
\,\Theta_\lambda \;=\; \lambda \, \Theta_\lambda
\eeq
involving the radial and angular operators defined by
\begin{eqnarray} \label{radial}
{\cal{R}}_\omega &=& \, -\frac{\partial}{\partial r}
\Delta \frac{\partial}{\partial r}
- \frac{1}{\Delta}((r^{2}+a^{2})\omega +ak)^{2} \\
{\cal{A}}_\omega &=& \label{angular}
-\frac{\partial}{\partial \cos \vartheta}\: \sin^2 \vartheta\:
\frac{\partial}{\partial \cos \vartheta}
+\frac{1}{\sin^2 \vartheta}(a\omega \sin^2 \vartheta + k )^{2} \:.
\end{eqnarray}
The separation constant $\lambda$ is an eigenvalue of the
angular operator~${\cal{A}}_\omega$. We denote the set of eigenvalues
of the {\em{spheroidal wave operator}}~${\cal{A}}_\omega$ by~$\{\lambda_n(\omega), \,n\in \N\}$
and normalize the corresponding eigenfunctions~$\Theta_{n,\omega}$
with respect to the~$L^2$-norm,
\beq \label{angnorm}
\| \Theta_{n,\omega} \|_{L^2(S^2)} \;=\; 1\:.
\eeq

Setting
\beq \label{phidef}
\phi(r) \;=\; \sqrt{r^2+a^2}\: R(r) \:,
\eeq
the radial equation can be written in Schr\"odinger-type form~\cite[Section~5]{FKSY}
\begin{equation}
\left(-\frac{\partial^2}{\partial u^2} + V(u) \right) \phi(u) \;=\; 0
\label{schroedinger}
\end{equation}
with the potential
\begin{equation} \label{5V}
V(u) \;=\; -\left( \omega + \frac{ak}{r^2+a^2} \right)^2 \:+\:
\frac{\lambda\:\Delta}{(r^2+a^2)^2} \:+\: \frac{1}{\sqrt{r^2+a^2}}\; \partial_u^2 \sqrt{r^2+a^2} \,.
\end{equation}

In these terms, the integral representation~\cite[Theorem~7.1]{FKSY2} of
the solution of the Cauchy problem~(\ref{cauchy}) becomes
\begin{equation}\label{intrep}
\Psi(t,r,\vartheta) \;=\; \frac{1}{2 \pi}
\sum_{n \in \sN} \int_{-\infty}^\infty \frac{d\omega}{\omega
\Omega}\:e^{-i \omega t} \sum_{a,b=1}^2 t^{\omega n}_{ab}\:
\Psi_a^{\omega n}(r,\vartheta)\; \bra \Psi_b^{\omega n}, \Psi_0 \ket\:,
\end{equation}
where the sums and the integrals converge
in~$L^2_{\mbox{\scriptsize{loc}}}$. Here
the coefficients $t^{\omega n}_{ab}$ are given by \beq
\label{tabdef} t^{\omega n}_{11} \;=\; 1 +
{\mbox{\rm{Re}}}\,\frac{\alpha}{\beta} \:,\qquad t^{\omega n}_{12} \;=\;
t^{\omega n}_{21} \;=\; -{\mbox{\rm{Im}}}\, \frac{\alpha}{\beta} \:,\qquad
t^{\omega n}_{22} \;=\; 1 - {\mbox{\rm{Re}}}\, \frac{\alpha}{\beta}\:, \eeq
and the complex coefficients $\alpha$ and $\beta$, referred to as
transmission coefficients, are defined by
\beq \label{trans} \grave{\phi}
\;=\; \alpha\: \acute{\phi} + \beta\: \overline{\acute{\phi}} \:,
\eeq
where~${\acute{\phi}}(u)$ and~${\grave{\phi}}(u)$ are the Jost solutions
of the radial equation having the asymptotics
\begin{eqnarray}
\lim_{u \to -\infty} e^{-i \Omega u} \:\acute{\phi}(u) &=& 1 \:,\spc
\lim_{u \to -\infty} \left(e^{-i \Omega u} \:\acute{\phi}(u) \right)' \;=\; 0  \label{abc1} \\
\lim_{u \to \infty} e^{i \omega u} \:\grave{\phi}(u) &=& 1 \:,\spc\;\;\;\;\,
\lim_{u \to \infty} \left(e^{i \omega u} \:\grave{\phi}(u) \right)' \;=\; 0  \label{abc2}
\end{eqnarray}
and
\[ \Omega \;=\; \omega - \omega_0 \spc {\mbox{where}} \spc
\omega_0 \;:=\; -\frac{ak}{r_1^2+a^2}\:. \]
(For ease in notation, we have omitted the $\omega$-dependence of~$\acute{\phi}$ and~$\grave{\phi}$.)
Finally, the functions
$\Psi_a^{\omega n}(r,\vartheta)$, $a=1,2$, are the solutions of
the separated wave equation (\ref{sepwave}) for fixed~$\omega$
and~$\lambda=\lambda_n(\omega)$, corresponding to the real-valued
fundamental solutions of the radial equation given by
\beq \label{phi12}
\phi_1 \;=\; {\mbox{Re}}\,{\acute{\phi}}\;,\qquad
\phi_2 \;=\; {\mbox{Im}}\, {\acute{\phi}}\:.
\eeq

\section{Single Mode Estimates}
\setcounter{equation}{0}
In this section we analyze a single summand in the integral representation~(\ref{intrep}).
We begin with the following lemma.
\begin{Lemma} \label{lemma1}
For any~$n \in \N$, the following hold:
\begin{description}
\item[(i)] For all~$\omega \in \R \setminus \{0, \omega_0\}$, $|\Psi^{\omega n}_a(r,\vartheta)| \leq c(n,r)\, (1+|\omega|)$, where the
constant is locally uniform in~$r$.
\item[(ii)] For all~$\omega \in \R \setminus \{0, \omega_0\}$, $|t^{\omega n}_{ab}|\leq c(n)\, |\Omega|$.
\item[(iii)] The functions~$\displaystyle \frac{\bra \Psi^{n\omega}_b, \Psi_0 \ket}{\omega}$ are
bounded in~$\omega$ and have rapid decay as~$\omega \rightarrow \pm \infty$.
\end{description}
\end{Lemma}
{\Proof} The statement~{\bf{(i)}} is an immediate consequence of the WKB-estimates
in~\cite[Proposition~6.5]{FKSY}, which show that for large~$|\omega|$, the
fundamental solutions~$\acute{\phi}$ go over to plane waves with amplitude one.

For any~$\omega$, the coefficients~$t^{\omega n}_{ab}$ are related to the
fundamental solutions~$\acute{\phi}$ and~$\grave{\phi}$ of the radial equation
and their Wronskian~$w(\acute{\phi}, \grave{\phi})$ by (see \cite[Lemma~5.1]{FKSY2})
\begin{eqnarray}
g(u,u') &=& \frac{\acute{\phi}(u)\: \grave{\phi}(u')}{w(\acute{\phi}, \grave{\phi})} 
\label{gdef} \\
{\mbox{Im}}\, g(u,u') &=& -\frac{1}{2 \Omega}\:
\sum_{a,b=1}^2 t^{\omega n}_{ab} \: \phi_a(u)\: \phi_b(u')\;. \label{tabrel}
\end{eqnarray}
Using the asymptotics near~$u=-\infty$
\[ \phi_1(u) \sim \cos(\Omega u)\:,\qquad \phi_2(u') \sim \sin(\Omega u')\:, \]
we evaluate~(\ref{tabrel}) for~$u, u'$ near~$-\infty$ to obtain the estimate
\[ |t^{\omega n}_{ab}| \;\leq\; 2 |\Omega|\: \sup_{u,u'} |g(u,u')|\:. \]
The causality argument in~\cite[Section~7]{FKSY2} implied that the Wronskian has no zeros,
and thus~$g(u,u')$ is bounded for~$\omega$ in any compact set.
For large~$|\omega|$, we can again use the WKB-estimates in~\cite[Proposition~6.5]{FKSY}
to conclude that~$g(u,u')$ decays like~$1/|\omega|$. This proves~{\bf{(ii)}}.

To prove~{\bf{(iii)}}, we first note that the energy inner product involves
a factor of~$\omega$ (see~\cite[eq.~(2.14)]{FKSY}). Therefore, the
quotient~$\bra \Psi^{\omega n}_a, \Psi_0 \ket/\omega$ is bounded near~$\omega=0$.
It remains to show that the inner product~$\bra \Psi^{\omega n}_a, \Psi_0 \ket$
has rapid decay in~$\omega$. We write the radial ODE~(\ref{schroedinger}) as
\[ \omega^2\, \acute{\phi} \;=\; (-\partial_u^2 + (V+\omega^2))\, \acute{\phi} \:. \]
Iterating this relation gives
\[ \omega^{2l}\, \acute{\phi} \;=\; (-\partial_u^2 + (V+\omega^2))^l\, \acute{\phi} \:. \]
We thus obtain for suitable functions~$F$ and~$G$,
which are independent of~$\omega$, the formula
\[ \omega^{2l}\, \acute{\Psi}^{\omega n} \;=\; (-\partial_u^2 + \omega F + G)^l\,
\acute{\Psi}^{\omega n}\:, \]
where~$\acute{\Psi}^{\omega n} = (\acute{\Phi}^{\omega n}, i \partial_t
\acute{\Phi}^{\omega n})$ and~$\acute{\Phi}^{\omega n}$ is defined by~(\ref{phidef},
\ref{Phisep}). Hence
\begin{eqnarray}
\bra \acute{\Psi}^{\omega n}, \Psi_0 \ket &=& \frac{1}{\omega^{2l}}
\: \bra (-\partial_u^2 + \omega F + G)^{l}\, \acute{\Psi}^{\omega n}, \Psi_0 \ket \nonumber \\
&=& \frac{1}{\omega^{2l}} \: \bra \acute{\Psi}^{\omega n}, ((-\partial_u^2 + \omega F + G)^*)^{l}\, \Psi_0 \ket\:, \label{nonum}
\end{eqnarray}
where star denotes the formal adjoint obtained by partial integration.
Writing the function~$((-\partial_u^2 + \omega F + G)^*)^l \Psi_0$ as a polynomial in~$\omega$,
each coefficient is again smooth with compact support.
Hence for large~$|\omega|$,
\[ \left| \bra \acute{\Psi}^{\omega n}, ((-\partial_u^2 + \omega F + G)^*)^{l} \Psi_0 \ket \right|
\leq\; C\, |\omega|^{l+1}\:, \]
where the constant~$C$ depends only on~$\Psi_0$ and~$l$. Combining this with~(\ref{nonum})
and using that~$l$ is arbitrary proves~{\bf{(iii)}}.
\QED
This lemma shows in particular that the integral
\[ \int_{-\infty}^\infty \frac{d\omega}{\omega
\Omega}\:e^{-i \omega t} \sum_{a,b=1}^2 t^{\omega n}_{ab}\:
\Psi_a^{\omega n}(r,\vartheta)\; \bra \Psi_b^{\omega n}, \Psi_0 \ket \]
which appears in our integral representation~(\ref{intrep}) is absolutely convergent.

Our next goal is to introduce a functional calculus for a single angular momentum mode.
In what follows, $C^\infty$ denotes the set of all bounded smooth functions,
all of whose derivatives are also bounded.
\begin{Lemma} \label{lemma2}
For any~$n \in \N$ and~$f \in C^\infty(\R)$, the function
\beq \label{fndef}
(f(H)_n \Psi_0)(u, \vartheta) \;:=\; \frac{1}{2 \pi} \int_{-\infty}^\infty \frac{d\omega}{\omega
\Omega}\:f(\omega) \sum_{a,b=1}^2 t^{\omega n}_{ab}\:
\Psi_a^{\omega n}(u,\vartheta)\; \bra \Psi_b^{\omega n}, \Psi_0 \ket
\eeq
lies in~$L^2(\R \times S^2, \,r^2 \,dr\, d\cos \vartheta)^2$.
\end{Lemma}
{\Proof} In view of Lemma~\ref{lemma1}, the function~$(f(H)_n \Psi_0)$ is bounded
on any compact set. Hence we only need to consider its behavior
near~$u=\pm \infty$. In the region~$u \ll 0$,
the proof of~\cite[Theorem~3.1]{FKSY2} shows that $\acute{\phi}$ can be uniformly
approximated by a plane wave, i.e.\ there are positive constants~$c$ and~$\gamma$
such that for all~$\omega \in \R$ and all~$u<u_0 \ll 0$,
\beq \label{aprep}
\acute{\phi}(u) \;=\; e^{i \Omega u} + E_\omega(u) \qquad {\mbox{with}} \qquad
|E_\omega(u)| \;\leq\; c\: e^{\gamma u} \:.
\eeq
The function~$E_\omega$ gives rise to a contribution in~(\ref{fndef}) which decays
exponentially as~$u \rightarrow -\infty$. On the other hand, according to~(\ref{phidef})
and Lemma~\ref{lemma1}, the plane wave gives rise to terms of the form
\beq \label{term}
(r^2+a^2)^{-\frac{1}{2}}\: \int_{-\infty}^\infty Y(\omega, \vartheta)\: e^{\pm i \omega u}
\, d\omega \:,
\eeq
where $Y(.,\vartheta)$ is in~$L^2$ with its~$L^2$-norm bounded uniformly in~$\vartheta$.
Plancherel's theorem yields that~(\ref{term}) is in~$L^2((-\infty, u_0) \times S^2)^2$.

For the asymptotics $u>u_1 \gg 0$, it is useful to introduce the quantity
\[ \omega_1 \;=\; \frac{1}{16 a k}\:. \]
We first bring the sum in~(\ref{fndef}) into a convenient
form. Using~(\ref{gdef}) and~(\ref{tabrel}), we know that
\begin{eqnarray}
\sum_{a,b=1}^2 t^{\omega n}_{ab} \: \phi^a(u')\: \phi^b(u)
&=& -2 \Omega \,{\mbox{Im}}\: \frac{\acute{\phi}(u')\: \grave{\phi}(u)}{w(\acute{\phi}, \grave{\phi})}
\nonumber \\
&=& i \Omega \left(
\frac{\acute{\phi}(u')}{w(\acute{\phi}, \grave{\phi})}\:\grave{\phi}(u)
- \overline{ \frac{\acute{\phi}(u')}{w(\acute{\phi}, \grave{\phi})}\:\grave{\phi}(u) }\: \right).
\label{tabdec}
\end{eqnarray}
We only consider the term involving~$\grave{\phi}$, because the term involving~$\overline{\grave{\phi}}$
can be treated similarly. To bound the quotient~$\acute{\phi}(u')/w(\acute{\phi}, \grave{\phi})$,
we note that for large~$|\omega|$, the WKB-estimates of~\cite[Section~6]{FKSY} yield
\beq \label{whe}
\left|\frac{\acute{\phi}(u')}{w(\acute{\phi}, \grave{\phi})} \right| \;\leq\; \frac{c}{|\omega|}\: .
\eeq
Since the Wronskian has no zeros, by increasing~$c$ we can always arrange that this
inequality holds on a compact set outside the origin.
We choose~$c$ such that~(\ref{whe}) holds for all~$\omega$ with~$|\omega| > \omega_1/2$.
On the other hand, if~$\omega$ is small, 
from~\cite[Theorem~3.5]{FKSY2} together with the argument before Proposition~5.2 in~\cite{FKSY2} we know
that~$w(\acute{\phi}, \omega^{\mu-\frac{1}{2}} \grave{\phi})$ is bounded away from zero, where
\[ \mu \;=\; \mu(\omega) \;=\; \sqrt{\lambda_n(\omega) - 2ak \omega + \frac{1}{4}}
\;>\; \frac{1}{3} \:, \]
and the last inequality follows from the fact that~$\lambda_n \geq 0$ and~$|\omega|<\omega_1 = (16ak)^{-1}$.
Hence
\beq \label{wle}
\left|\frac{\acute{\phi}(u')}{w(\acute{\phi}, \grave{\phi})} \right| \;\leq\; c\:|\omega|^{\mu-\frac{1}{2}}
\spc {\mbox{if~$|\omega| \leq \omega_1$}} \:.
\eeq

We choose a cutoff function~$\eta \in C^\infty_0([-\omega_1, \omega_1])$, which on the
interval~$[-\omega_1/2, \omega_1/2]$ is is identically equal to one. We split up~$f$
in the form
\[ f \;=\; f_H + f_L \qquad {\mbox{with}} \qquad f_L \;=\; \eta\, f\:,\quad f_H \;=\; (1-\eta)\, f\:, \]
and consider the corresponding ``high- and low-energy contributions'' to~(\ref{fndef}) separately.
To estimate the high-energy contribution, we write~(\ref{fndef}) (omitting the second
term in~(\ref{tabdec})) as
\beq \label{ghe2}
(r^2+a^2)^{-\frac{1}{2}} \int_{-\infty}^\infty X(\omega, \vartheta)\: \grave{\phi}(u)\, d\omega \:,
\eeq
where
\[ X(\omega, \vartheta) \;:=\; i \Omega\: \frac{\bra \acute{\Psi}^{\omega n}, \Psi_0 \ket}
{w(\acute{\phi}, \grave{\phi})} \: \Theta_{n, \omega}(\vartheta) \:. \]
In view of~(\ref{whe}) and Lemma~\ref{lemma1},
the function~$X$ is bounded and has rapid decay.
Furthermore, differentiating with respect to~$\omega$,
we can use that both~$\acute{\phi}$ and the angular eigenfunction
are smooth in~$\omega$. Furthermore, the Jost expansion~\cite[eq.~(3.18, 3.19)]{FKSY2}
shows that $|\partial_\omega \grave{\phi}| < c/|\omega|$ for small~$|\omega|$.
For large~$|\omega|$, we can again use the WKB estimates in~\cite{FKSY} to get rapid decay
of~$X'$. Hence
\beq \label{Xpes}
|\partial_\omega X(\omega, \vartheta)| \;\leq\; \frac{c}{|\omega|} \qquad
{\mbox{and}} \qquad {\mbox{$\partial_\omega X(\omega, \vartheta)$ has rapid decay}}\:.
\eeq

Note that the factor~$(r^2+a^2)^{-1/2}$ in~(\ref{ghe2})
is taken care of by the integration
measure~$r^2 \,dr \,d(\cos \vartheta)$, and thus we may omit this factor
and work instead with the integration measure~$du$.
We now decompose~$\grave{\phi}$ similar to~(\ref{aprep})
in the form~\cite[Lemma~3.3]{FKSY2},
\beq \label{agrep}
\grave{\phi}(u) \;=\; e^{-i \omega u} + E_\omega(u) \qquad {\mbox{with}} \qquad
|E_\omega(u)| \;\leq\; \frac{c}{u} \:.
\eeq
The contribution of~$E_\omega(u)$  to~(\ref{ghe2}) decays like~$1/u$ and
is thus in~$L^2$. The contribution of the plane wave~$e^{-i \omega u}$
to~(\ref{ghe2}) is in~$L^2$ according to Plancherel's theorem.

It remains to estimate the low-energy $f_L$-contribution to~(\ref{fndef}).
We first note that, according to~\cite[Lemma~3.6]{FKSY2},
the functions~$\grave{\phi}$ are well-approximated near~$\omega=0$ by
Hankel functions, i.e.\ there is a constant~$c$ such that for all~$u>u_1 \gg 0$ and all~$\omega
\in (-\omega_1, \omega_1)$,
\beq \label{Edef2}
\grave{\phi}(u) \;=\; H(\omega u) + E_\omega(u)
\qquad {\mbox{with}} \qquad
|E_\omega(u)| \;\leq\; \frac{c \omega^{-\mu+\frac{1}{2}}}{u} \:.
\eeq
Here~$H(x)$ is a Hankel function, which for~$|x|<1$ has the asymptotics
\beq \label{Has1}
H(x) \;=\; c\, |x|^{-\mu+\frac{1}{2}} \:(1+ {\mathcal{O}}(x)) \:,
\eeq
whereas for~$|x|>1$ we can write it as
\beq \label{Has2}
H(x) \;=\; e^{-i x} \, h(x) \qquad {\mbox{with}} \qquad
\left| \frac{d^n}{dx^n} h(x) \right| \;\leq\; \frac{c_n}{|x|^n}, \quad
n \geq 0 \:.
\eeq
Using~(\ref{Edef2}, \ref{tabdec}, \ref{wle}), we obtain for all~$u>u_1$ the estimate
\begin{eqnarray}
\lefteqn{ \left| \int_{-\omega_1}^{\omega_1} \frac{d\omega}{\omega \Omega}\:f_L\: 
\sum_{a,b=1}^2 t^{\omega n}_{ab}\:
\Psi_a^{\omega n}(r,\vartheta)\; \bra \Psi^b_{k \omega n}, \Psi_0 \ket \right| } \nonumber \\
&\leq& C (r^2+a^2)^{-\frac{1}{2}} \left|
\int_{-\omega_1}^{\omega_1} X(\omega, \vartheta)\: \omega^{\mu-\frac{1}{2}}
\left( H(\omega u) + E_\omega(u) \right) d\omega \right| . \label{f}
\end{eqnarray}
According to~(\ref{Edef2}), the contribution of~$E_\omega$ to~(\ref{f}) decays at infinity like~$1/u$, and thus is in~$L^2$.
To estimate the contribution by the Hankel function, we split up the integral as follows,
\[ \int_{-\omega_1}^{\omega_1} \;=\; \int_{-1/u}^{1/u} + \int_{[-\omega_1,\omega_1]
\setminus [-1/u, 1/u]} . \]
The first integral is estimated using the asymptotics~(\ref{Has1}),
\[ \int_{-1/u}^{1/u} |X(\omega, \vartheta)|\: |\omega|^{\mu-\frac{1}{2}}\,
|H(\omega u)|\:d\omega \;\leq\; \frac{c \|X\|_\infty}{u^{\mu
+\frac{1}{2}}} \:. \]
In the second integral we use the asymptotics~(\ref{Has2}) and integrate by parts,
\begin{eqnarray}
\lefteqn{ \int_{[-\omega_1,\omega_1] \setminus [-1/u, 1/u]}
X(\omega, \vartheta)\: \omega^{\mu-\frac{1}{2}}\,
H(\omega u) \:d\omega } \nonumber \\
&=&  u^{-\mu-\frac{1}{2}}
\int_{[-\omega_1 u,\omega_1 u] \setminus [-1, 1]} X\!\left(\frac{x}{u}, \vartheta \right)\: 
e^{-ix} \:x^{\mu-\frac{1}{2}} \,h(x) \:dx \label{ex0} \\
&=& u^{-\mu-\frac{1}{2}}
\int_{[-\omega_1 u,\omega_1 u] \setminus [-1, 1]} X\!\left(\frac{x}{u}, \vartheta \right)\:
i \frac{d}{dx} \left(e^{-ix} \right)
x^{\mu-\frac{1}{2}} \,h(x) \:dx \nonumber \\
&=& i u^{-\mu-\frac{1}{2}} \left. X\!\left(\frac{x}{u}, \vartheta \right)\: 
e^{-ix}\: x^{\mu-\frac{1}{2}} \,h(x) \right|_{x=-\omega_1 u}^{x=\omega_l u} \label{exm1} \\
&& -i u^{-\mu-\frac{1}{2}} \left. X\!\left(\frac{x}{u}, \vartheta \right)\: 
e^{-ix}\: x^{\mu-\frac{1}{2}} \,h(x) \right|_{x=-1}^{x=1} \label{ex1} \\
&&-i  u^{-\mu-\frac{1}{2}} \int_{[-\omega_1 u,\omega_1 u] \setminus [-1, 1]} D_1 X \!\left(\frac{x}{u}, \vartheta \right)\:\frac{1}{u}\: 
e^{-ix} \,x^{\mu-\frac{1}{2}}\,h(x) \:dx \label{ex2} \\
&&-i u^{-\mu-\frac{1}{2}}
\int_{[-\omega_1 u,\omega_1 u] \setminus [-1, 1]} X\!\left(\frac{x}{u}, \vartheta \right)\:
e^{-ix}\: \frac{d}{dx} \left( x^{\mu-\frac{1}{2}} \,h(x) \right) dx\:. \label{ex3}
\end{eqnarray}
Since~$X$ is a bounded function, the terms~(\ref{exm1}, \ref{ex1}) are of the order~${\mathcal{O}}(u^{-\mu-\frac{1}{2}})$
and are thus in~$L^2$. To estimate the term~(\ref{ex2}), we transform back to the integration variable~$\omega$,
\[ (\ref{ex2}) \;=\; -\frac{i}{u}
\int_{[-\omega_1, \omega_1] \setminus [-1/u, 1/u]} \partial_\omega X\!\left(\omega, \vartheta \right)\:
\omega^{\mu-\frac{1}{2}}\,e^{-i\omega u} \,h(\omega u) \:d\omega \:, \]
and since~$h$ is a bounded function and~$\partial_\omega X$ has rapid decay
according to~(\ref{Xpes}), the integral is uniformly bounded
in~$u$ and~$\vartheta$. Thus~(\ref{ex2}) is also in~$L^2$.
The last term~(\ref{ex3}) is similar to~(\ref{ex0}), except that,
in view of~(\ref{Has2}), the $x$-derivative has improved the decay rate of~(\ref{ex3})
at~$x=\pm \infty$ by a factor of~$1/x$. Iterating the above procedure~(\ref{ex0}--\ref{ex3}),
the last term becomes
\[ u^{-\mu-\frac{1}{2}}
\int_{[-\omega_1 u,\omega_1 u] \setminus [-1, 1]} X\!\left(\frac{x}{u}, \vartheta \right)\:
e^{-ix}\: \left\{ \left(-i \frac{d}{dx} \right)^l \!\!\left( x^{\mu-\frac{1}{2}} \,h(x) \right) \right\} dx\:. \]
After choosing~$l$ sufficiently large and using~(\ref{Has2}), 
the curly bracket term is in~$L^1(\R, dx)$. Since~$X$ is a bounded function, the integral is
bounded uniformly in~$u$. Thus the whole expression is bounded
by a constant times~$u^{-\mu-\frac{1}{2}}$, and this is an $L^2$-function.
\QED

By applying the last lemma with initial data~$H^s \Psi_0$, we obtain that all Sobolev
norms are also bounded for each mode; namely, we have the following corollary.
\begin{Corollary} \label{cor1}
For every~$n, s \in \N$ and any~$f \in C^\infty(\R)$,
the function~$(f(H)_n \Psi_0)$ lies in $H^{s,2}((r_1, \infty) \times S^2, \,r^2 dr\, d\cos \vartheta)$.
Moreover, there are constants~$c>0$ and~$l \in \N$ (depending on~$n$, $s$
and~$\Psi_0$, but independent of~$f$) such that
\[ \|f(H)_n \Psi_0\|_{H^{s,2}}
\;\leq\; c\:\|f\|_{C^l} \:. \]
\end{Corollary}
{\Proof} Since in the proof of the last lemma we integrated by parts a finite
number of times and used only the boundedness of~$f$ and its derivatives, it is
obvious that there are constants~$c$ and~$l$ such that
\beq \label{l2bound}
\|f(H)_n \Psi_0\|_{L^2} \;\leq\; c\:\|f\|_{C^l}\:.
\eeq
Using~(\ref{Hdef}, \ref{Adef}) we see that the operator~$H^2$ is uniformly
elliptic. Hence
\beq \label{l3bound}
\|f(H)_n \Psi_0\|_{H^{s,2}} \;\leq\; c(n)\, \sum_{k=0}^s
 \|H^{2k} f(H)_n \Psi_0\|_{L^2} \:.
\eeq
From Lemma~\ref{lemma1} and the fact that the spatial derivatives of~$\Psi_a^{\omega n}$
can be bounded using the radial and angular ODEs by powers of~$\omega$,
we may differentiate through~(\ref{fndef}) using Lebesgue's dominated convergence theorem
to obtain
\begin{eqnarray*}
\lefteqn{ H (f(H)_n \Psi_0)(u, \vartheta) \;=\; \int_{-\infty}^\infty \frac{d\omega}{\omega
\Omega}\:f(\omega) \sum_{a,b=1}^2 t^{\omega n}_{ab}\:
(H \Psi_a^{\omega n}(u,\vartheta))\; \bra \Psi_b^{\omega n}, \Psi_0 \ket } \\
&=& \int_{-\infty}^\infty \frac{d\omega}{\omega
\Omega}\:f(\omega) \sum_{a,b=1}^2 t^{\omega n}_{ab}\:
(\omega \Psi_a^{\omega n}(u,\vartheta))\; \bra \Psi_b^{\omega n}, \Psi_0 \ket \\
&=& \int_{-\infty}^\infty \frac{d\omega}{\omega
\Omega}\:f(\omega) \sum_{a,b=1}^2 t^{\omega n}_{ab}\:
\Psi_a^{\omega n}(u,\vartheta)\; \bra H \Psi_b^{\omega n}, \Psi_0 \ket \\
&=& \int_{-\infty}^\infty \frac{d\omega}{\omega
\Omega}\:f(\omega) \sum_{a,b=1}^2 t^{\omega n}_{ab}\:
\Psi_a^{\omega n}(u,\vartheta)\; \bra \Psi_b^{\omega n}, H \Psi_0 \ket
\;=\; (f(H)_n (H \Psi_0))(u, \vartheta)\:,
\end{eqnarray*}
and iterating this process yields that
\[ \| H^{2k} f(H)_n \Psi_0\|_{L^2} \;=\; \|f(H)_n (H^{2k} \Psi_0) \|_{L^2} \:. \]
Substituting this identity in~(\ref{l3bound}) and
applying~(\ref{l2bound}) gives the desired inequality.
\QED

We will also need the following~$L^1$-estimate for the radial solutions.
\begin{Lemma} \label{lemmal1}
For any~$\varepsilon>0$ and~$n \in \N$ there is a constant~$c=c(\varepsilon,n)$ such that for
all~$f \in C^\infty_0(\R \setminus (-\varepsilon, \varepsilon))$
the following inequality holds,
\beq \label{l1es}
\left\| \int_{\R} f(\omega) \,\acute{\phi}^{\omega n}(u)\, d\omega \right\|^2_{L^1(\R, du)}
\;\leq\; c\, \|f\|_{H^{1,1}(\R, d\omega)}\, \|f\|_{H^{2,1}(\R, d\omega)}\:.
\eeq
\end{Lemma}
{\Proof} Near the event horizon, we can again use the asymptotics~(\ref{aprep})
and bound the error term by
\[ \int_{-\infty}^{u_0} \left| \int_Q f(\omega) \,E_{\omega}(u) \,d\omega \right| du
\;\leq\; c \int_{-\infty}^{u_0} du \int_Q |f(\omega)| e^{\gamma u} \,d\omega
\;\leq\; \frac{c e^{\gamma u_0}}{\gamma}\|f\|_{L^1(Q)} \:, \]
where~$Q=\R \setminus (-\varepsilon, \varepsilon)$.
For the plane wave, we integrate by parts,
\[ e^{-i \omega_0 u} \int_Q f(\omega) \,e^{i \omega u}\, d\omega
\;=\; \frac{i}{u} \:e^{-i \omega_0 u} \int_Q f'(\omega) \,e^{i \omega u}\, d\omega
\;=\; -\frac{1}{u^2} \:e^{-i \omega_0 u} \int_Q f''(\omega)\, e^{i \omega u}\, d\omega\:, \]
to obtain the two pointwise bounds
\[ \left| \int_Q f(\omega) \,e^{i \Omega u} \,d\omega \right|
\;\leq\; \frac{c\, \|f\|_{H^{1,1}(Q)}}{|u|}, \frac{c\, \|f\|_{H^{2,1}(Q)}}{u^2} \]
and thus
\[ \left| \int_Q f(\omega) \,e^{i \Omega u} \,d\omega \right|
\;\leq\; \frac{c}{|u|^{\frac{3}{2}}}\: \sqrt{ \|f\|_{H^{1,1}(Q)} \, \|f\|_{H^{2,1}(Q)}} \:. \]
Since~$|u|^{-\frac{3}{2}} \in L^1((-\infty, u_0])$, we obtain the desired inequality~(\ref{l1es}).

Near infinity we use the asymptotics~(\ref{agrep}). The plane wave term can be treated
exactly as before. Using the explicit integral formulas~\cite[eqns~(3.16, 3.18, 3.19)]{FKSY2},
we can write the error term~$E_\omega$ in the form
\[ E_\omega(u) \;=\; \frac{h(\omega)\, e^{-i \omega u}}{u}\:+\: {\mathcal{O}}(u^{-2}) \]
with a smooth function~$h$. Again integrating by parts,
\[ \frac{1}{u} \int_Q (fh)(\omega) e^{i \omega u} d\omega \;=\;
\frac{i}{u^2} \int_Q (fh)'(\omega) e^{i \omega u} d\omega\:, \]
we see that all the resulting terms are in~$L^1([u_1, \infty), du)$.

On the remaining compact interval~$[u_0, u_1]$, we can simply bound the~$L^1$-norm by
$(u_1-u_0)$ times the sup-norm.
\QED

\section{Functional Calculus for a Finite Number of Modes}
\setcounter{equation}{0}
We begin this section with the following orthogonality relations for
any~$\Psi_0 \in C^\infty_0$.
\begin{Prp} \label{lemma31}
Suppose that~$f, g \in C^{\infty}(\R)$ with~$0 \not \in {\mbox{\rm{supp}}}\,f$
or~$0 \not \in {\mbox{\rm{supp}}}\,g$.
\begin{description}
\item[(i)] If~${\mbox{\rm{dist}}}({\mbox{\rm{supp}}}\, f, {\mbox{\rm{supp}}}\,g)>0$,
\[ \bra f(H)_n \Psi_0, g(H)_{n} \Psi_0 \ket \;=\; 0 \spc {\mbox{for all $n \in \N$}}\:.\]
\item[(ii)] $\spc\spc\qquad\; \bra f(H)_n \Psi_0, g(H)_{n'} \Psi_0 \ket \;=\; 0 \spc {\mbox{for all $n, n' \in \N$, $n \neq n'$}}$.
\end{description}
\end{Prp}
For the proof of this proposition we will need the following lemma.
\begin{Lemma} \label{lemmaangular}
For any~$n, n' \in \N$
there is a locally bounded function~$\alpha(\omega, \omega')$
such that for all~$\omega, \omega' \in \R$ the
angular eigenfunctions~$\Theta_{n,\omega}$ of the operator~${\mathcal{A}}_\omega$ satisfy
following identity
\beq \label{angex}
\langle \Theta_{n,\omega}, \Theta_{n',\omega'} \rangle_{L^2(S^2)}
\;=\; (\omega-\omega') \; \alpha(\omega, \omega') \:,
\eeq
where
\begin{eqnarray}
\alpha(\omega, \omega') &=&
\frac{2 a k}{\lambda_n(\omega) - \lambda_{n'}(\omega')}
 \, \langle \Theta_{n,\omega}, \Theta_{n',\omega'} \rangle_{L^2(S^2)} \nonumber \\
&&+\frac{a^2\, (\omega+\omega')}{\lambda_n(\omega) - \lambda_{n'}(\omega')} \, 
\langle \Theta_{n,\omega}, \sin^2 \vartheta\,\Theta_{n',\omega'} \rangle_{L^2(S^2)} \:.
\label{alphadef}
\end{eqnarray}
The function~$\alpha$ and its derivatives are polynomially bounded.
\end{Lemma}
{\Proof} Using the angular equation in~(\ref{sepwave}) we have
\begin{eqnarray*}
(\lambda_n(\omega) - \lambda_{n'}(\omega') \:\langle \Theta_{n, \omega}, \Theta_{n', \omega'} \rangle_{L^2(S^2)}
&=& \langle {\mathcal{A}}_\omega \Theta_{n, \omega}, \Theta_{n', \omega'} \rangle_{L^2(S^2)}
- \langle \Theta_{n, \omega}, {\mathcal{A}}_{\omega'} \Theta_{n', \omega'} \rangle_{L^2(S^2)} \\
&=& \langle \Theta_{n, \omega}, ({\mathcal{A}}_\omega-{\mathcal{A}}_{\omega'})
\Theta_{n', \omega'} \rangle_{L^2(S^2)}
\end{eqnarray*}
and thus
\begin{eqnarray*}
\langle \Theta_{n, \omega}, \Theta_{n', \omega'} \rangle_{L^2(S^2)} &=&
\frac{1}{\lambda_n(\omega) - \lambda_{n'}(\omega')}
\:\langle \Theta_{n, \omega}, ({\mathcal{A}}_\omega-{\mathcal{A}}_{\omega'})
\Theta_{n', \omega'} \rangle_{L^2(S^2)}\:.
\end{eqnarray*}
Since
\[ {\mathcal{A}}_\omega-{\mathcal{A}}_{\omega'} \;=\;
a^2\, (\omega^2 - \omega'^2)\, \sin^2 \vartheta + 2 k a\, (\omega-\omega')
\;=\; (\omega-\omega') \left(  a^2\, (\omega+\omega')\, \sin^2 \vartheta + 2 a k \right) , \]
we obtain~(\ref{angex}).

Using the gap estimates~\cite[Theorem~1.3]{FS}, we can apply standard analytic
perturbation theory~\cite{Kato} to conclude that~$\alpha$ and its derivatives
are polynomially bounded.
\QED

\noindent
{\em{Proof of Proposition~\ref{lemma31}.}} Without loss of generality we can assume that~$0 \not \in {\mbox{\rm{supp}}}\,g$.
According to Lemma~\ref{lemma1} we can write the spectral representations as
\begin{eqnarray*}
(f(H)_n \Psi_0)(r, \vartheta) &=& \int_{-\infty}^\infty f(\omega) \:h^a(\omega)\,
\Psi_a^{\omega n}(r, \vartheta)\: d\omega \\
(g(H)_{n'} \Psi_0)(r, \vartheta) &=& \int_{-\infty}^\infty g(\omega') \:h^a(\omega')\,
\Psi_a^{\omega' n'}(r, \vartheta)\: d\omega' \:,
\end{eqnarray*}
where $h^a(\omega)$ are bounded functions with rapid decay;
for notational convenience we omitted the sum over~$a=1,2$.
We choose a function~$\eta \in C^\infty_0((-2,2))$ which is identically equal to one
on the interval~$[-1,1]$. For~$K>0$ we set~$\eta_K(u)=\eta(u/K)$. Then using Corollary~\ref{cor1}
and Lebesgue's dominated convergence theorem,
\begin{eqnarray}
\lefteqn{ \bra f(H)_n \Psi_0, g(H)_{n'} \Psi_0 \ket \;=\;
\lim_{K \rightarrow \infty} \bra \eta_K\, f(H)_n \Psi_0, g(H)_{n'} \Psi_0 \ket } \nonumber \\
&=& \lim_{K \rightarrow \infty} \int_{-\infty}^\infty \overline{f(\omega)}\: 
\overline{h^a(\omega)}\, d\omega \int_{-\infty}^\infty 
g(\omega')\, h^b(\omega')\, d\omega'\: \bra \eta_K \Psi_a^{\omega n}, \Psi_b^{\omega' n'} \ket\:.
\label{dint}
\end{eqnarray}

In case~{\bf{(i)}} we restrict attention to~$n=n'$ and set
\beq \label{deltadef}
\delta \;=\; {\mbox{dist}} \left( {\mbox{supp}}\, f,
{\mbox{supp}}\, g \right) \;>\; 0\:.
\eeq
Since~$H$ is symmetric with respect to the energy inner product,
\begin{eqnarray*}
(\omega - \omega') \,\bra \eta_K \Psi_a^{\omega n}, \Psi_b^{\omega' n} \ket
&=& \bra \eta_K H \Psi_a^{\omega n}, \Psi_b^{\omega' n} \ket
- \bra \eta_K \Psi_a^{\omega n}, H \Psi_b^{\omega' n} \ket \\
&=& -\bra [H, \eta_K] \Psi_a^{\omega n}, \Psi_b^{\omega' n} \ket \:.
\end{eqnarray*}
The commutator~$[H, \eta_K]$ is supported in~$[-2K, -K] \cup [K,2K]$.
We treat these two intervals separately. On the interval~$[K,2K]$,
we carry out the~$\omega'$-integral to obtain
\begin{eqnarray*}
\lefteqn{ \int_{-\infty}^\infty \overline{f(\omega)}\: 
\overline{h^a(\omega)}\, d\omega \int_{-\infty}^\infty 
g(\omega')\, h^b(\omega')\, d\omega' \, \frac{1}{\omega-\omega'}
\bra \chi_{[K,2K]}\, [H, \eta_K] \Psi_a^{\omega n}, \Psi_b^{\omega' n} \ket } \\
&=& \int_{-\infty}^\infty \overline{f(\omega)}\: 
\overline{h^a(\omega)}\:
\bra \chi_{[K,2K]}\, [H, \eta_K] \Psi_a^{\omega n}, g_\omega(H)_n \Psi_0 \ket \: d\omega
\end{eqnarray*}
where
\[ g_\omega(\omega') \;=\; \frac{g(\omega')}{\omega-\omega'}\:. \]
For sufficiently large~$K$, the energy density is positive, and
thus we may apply the Schwarz inequality to get
\begin{eqnarray*}
\lefteqn{ \left| \int_{-\infty}^\infty \overline{f(\omega)}\: 
\overline{h^a(\omega)}\, d\omega \int_{-\infty}^\infty 
g(\omega')\, h^b(\omega')\, d\omega' \, \frac{1}{\omega-\omega'}
\bra \chi_{[K,2K]}\, [H, \eta_K] \Psi_a^{\omega n}, \Psi_b^{\omega' n} \ket \right| } \\
&\leq& \int_{-\infty}^\infty |f(\omega)|\: |h^a(\omega)|\;
\left( \bra \chi_{[K,2K]}\, [H, \eta_K] \Psi_a^{\omega n},
\chi_{[K,2K]}\, [H, \eta_K] \Psi_a^{\omega n} \ket
\right)^{\frac{1}{2}} \\
&&\quad\;\: \spc\spc \times 
\left( \bra \chi_{[K,2K]} g_\omega(H)_n \Psi_0, \chi_{[K,2K]} g_\omega(H)_n \Psi_0  \ket \right)^{\frac{1}{2}} \, d\omega\:.
\end{eqnarray*}
Applying Corollary~\ref{cor1} together with the estimate (cf~(\ref{deltadef}))
\[ \|g_\omega\|_{C^l} \;\leq\; c(l)\: \frac{\|g\|_{C^l}}{\delta^{l+1}}\:, \]
the last factor in the above integral is bounded.
Since each derivative of~$\eta_K$ gives a factor~$1/K$, and the integration range
is of order~$K$, the first inner product can be bounded by a polynomial in~$\omega$
times~$1/K$. Using the fact that~$h^a$ has rapid decay, we conclude that the
whole expression tends to zero as~$K \rightarrow \infty$.

To complete the proof of~{\bf{(i)}}, it remains to show that the expression
\[ \int_{-\infty}^\infty \overline{f(\omega)}\: 
\overline{h^a(\omega)}\, d\omega \int_{-\infty}^\infty 
g(\omega')\, h^b(\omega')\, d\omega'\:\frac{1}{\omega-\omega'}\:
\bra \chi_{[-2K, -K]}\, [H, \eta_K] \Psi_a^{\omega n}, \Psi_b^{\omega' n} \ket \]
vanishes as~$K \rightarrow \infty$.
Using the asymptotic estimate~(\ref{aprep}) for~$\Psi_a^{\omega n}$ and~$\Psi_b^{\omega' n}$,
the exponential decay of the error term near the event horizon allows us to take the limit~$K \rightarrow \infty$
using Lebesgue's dominated convergence theorem to get zero for the contribution of the error
terms. Hence we may replace~$\Psi_a^{\omega n}$ and~$\Psi_b^{\omega' n}$ by
the plane waves in the asymptotic expansion~(\ref{aprep}). Substituting the identity
\[ e^{i (\Omega-\Omega') u} \;=\; -\frac{i}{\Omega-\Omega'}\: \frac{d}{du}
e^{i (\Omega-\Omega') u} \]
and integrating by parts, the $u$-derivative hits either the function~$\eta_K$ or
one of the coefficient functions. Since the derivatives of the coeffient functions
decay exponentially near the event horizon, the corresponding contributions
again vanish as~$K \rightarrow \infty$. We are thus left with the following term,
\[ \int_{-\infty}^\infty \overline{f(\omega)}\: 
\overline{h^a(\omega)}\, d\omega \int_{-\infty}^\infty 
g(\omega')\, h^b(\omega')\, d\omega'\:\:\frac{1}{(\omega-\omega')^2}\:
\bra \chi_{[-2K, -K]}\, [H, \eta'_K] \Psi_a^{\omega n}, \Psi_b^{\omega' n} \ket \:. \]
Since this involves at least two derivatives of~$\eta_K$, each one of which
gives a factor~$1/K$, the whole expression is of the order~$1/K$, and thus
tends to zero in the limit~$K \rightarrow \infty$. This completes the proof of~{\bf{(i)}}.

To prove~{\bf{(ii)}}, we choose a non-negative test function~$\zeta \in C^\infty_0([-2,2])$ which
is identically equal to one on~$[-1,1]$. For any~$\delta>0$ we
set~$\zeta_\delta(x) = \zeta(x/\delta)$. We return to~(\ref{dint}) before
taking the limit~$K \rightarrow \infty$ obtaining
\[ \bra \eta_K f(H)_n \Psi_0, g(H)_{n'} \Psi_0 \ket
\;=\; \int_{-\infty}^\infty \overline{f(\omega)}\: 
\overline{h^a(\omega)}\, d\omega \int_{-\infty}^\infty 
g(\omega')\, h^b(\omega')\, d\omega'\:
\bra \eta_K \Psi_a^{\omega n}, \Psi_b^{\omega' n'} \ket \:. \]
Inserting the identity $1=\zeta_\delta(\omega-\omega') + (1-\zeta_\delta(\omega-\omega'))$
gives
\begin{eqnarray}
\lefteqn{ \bra \eta_K f(H)_n \Psi_0, g(H)_{n'} \Psi_0 \ket } \nonumber \\
&=& \int_{-\infty}^\infty \overline{f(\omega)}\: 
\overline{h^a(\omega)}\, d\omega \int_{-\infty}^\infty 
g(\omega')\, h^b(\omega')\, d\omega'\:\zeta_\delta(\omega-\omega') \:
\bra \eta_K \Psi_a^{\omega n}, \Psi_b^{\omega' n'} \ket \label{ins1} \\
&&+ \int_{-\infty}^\infty \overline{f(\omega)}\: 
\overline{h^a(\omega)}\, d\omega \int_{-\infty}^\infty 
g(\omega')\, h^b(\omega')\, d\omega'\:(1-\zeta_\delta(\omega-\omega'))\:
 \bra \eta_K \Psi_a^{\omega n}, \Psi_b^{\omega' n'} \ket \:. \quad\;\; \label{ins2}
\end{eqnarray}
Noting that the integrand in~(\ref{ins2}) 
vanishes unless~$|\omega-\omega'|>\delta$, the argument in~{\bf{(i)}}
can be applied again, and hence this term goes to zero as~$K \rightarrow \infty$.

For the term~(\ref{ins1}) we proceed as follows. We write the energy inner product
in the form~\cite[eq.~(2.14)]{FKSY},
\begin{eqnarray}
\lefteqn{ \bra \eta_K \Psi_a^{\omega n}, \Psi_b^{\omega' n'} \ket \;=\;
\omega' \int_{r_1}^\infty dr \int_{-1}^1 d(\cos\vartheta)\, \eta_K(u) } \nonumber \\
&& \hspace*{-.5cm} \times \left[
\left( \frac{(r^2+a^2)^2}{\Delta} - a^2 \sin^2 \vartheta \right) (\omega + \omega')\,
\overline{\Phi_a^{\omega n}}
\:\Phi_b^{\omega' n'} + 2 a k \left(\frac{r^2+a^2}{\Delta}-1 \right) \overline{\Phi_a^{\omega n}}
\:\Phi_b^{\omega' n'} \right]. \;\;\quad \label{eiprep}
\end{eqnarray}
The term involving~$\sin^2 \vartheta$,
\[ -\omega' (\omega + \omega') \int_{r_1}^\infty dr \int_{-1}^1 d(\cos\vartheta)\, \eta_K(u)\,
a^2 \sin^2 \vartheta \:\overline{\Phi^a_{\omega,n}} \:\Phi_b^{\omega' n'} \:, \]
can be estimated using~(\ref{Phisep}, \ref{phidef}) and~(\ref{angnorm}) by
\[ |\omega' (\omega + \omega')|  \,a^2\, \sup_{u \in \R}
\left( \eta_K\, |\acute{\phi}_{\omega,n}|\:|\acute{\phi}^{\omega' n'}| \right)
\int_{r_1}^\infty \frac{dr}{r^2+a^2} \:, \]
which can clearly bounded by a polynomial~$P(\omega, \omega')$, uniformly in~$K$.
The corresponding contribution in~(\ref{ins1}) is estimated by
\begin{eqnarray*}
\lefteqn{ \int_{-\infty}^\infty |f h^a|(\omega)\, d\omega \int_{-\omega-\delta}^{\omega+\delta} 
|g h^b|(\omega')\, d\omega'\: \zeta_\delta(\omega-\omega') \:
P(\omega, \omega') } \\
&\leq& 2 \delta \int_{-\infty}^\infty |f h^a|(\omega)\,
\sup_{\omega' \in (\omega-\delta, \omega+\delta)} \left( 
|g h^b|(\omega')\, d\omega'\: \zeta_\delta(\omega-\omega') \:
P(\omega, \omega') \right) d\omega \:.
\end{eqnarray*}
Noting that, in view of Lemma~\ref{lemma1}, $f$ and~$g$ are bounded, whereas~$h^a$ and~$h^b$ 
have rapid decay, the integral is bounded independent of~$K$.
We conclude that the term involving~$\sin^2 \vartheta$ is of order~$\delta$, uniformly in~$K$.

The remaining terms in~(\ref{eiprep}) involve only radial coefficient functions.
Transforming to the Regge-Wheeler coordinate and the radial function~$\phi^a$,
we are left with
\beq \label{exthis}
\omega' \int_{-\infty}^\infty du\, \eta_K(u)
\left[ (\omega + \omega')\,
- 2 a k \:\frac{2 M r}{(r^2+a^2)^2} \right] 
\overline{\phi^a_{\omega, n}(u)} \:\phi_b^{\omega' n'}(u) \;
\langle \Theta_{n,\omega}, \Theta_{n',\omega'} \rangle_{L^2(S^2)} .
\eeq
To the angular inner product we
apply the identity from Lemma~\ref{lemmaangular}. Since~$|\omega-\omega'|<\delta$,
after choosing~$\delta$ sufficiently small
we may use the gap estimates from~\cite[Theorem~1.3]{FS} together with the
smooth dependence of the eigenvalues on~$\omega$ to conclude that the
function~$\lambda_n(\omega)-\lambda_{n'}(\omega')$ is smooth and bounded away from zero.
Thus we can write~(\ref{exthis}) in the form
\[ (\omega-\omega') \int_{-\infty}^\infty du\, \eta_K(u)\,
\overline{\phi^a_{\omega, n}(u)} \:\phi_b^{\omega' n'}(u) \;
\left( B^1(\omega, \omega') + B^2(\omega, \omega')\: \frac{r}{(r^2+a^2)^2} \right) \:, \]
where
\beq \label{Bdef}
B^1(\omega, \omega') \;=\; \omega' (\omega + \omega')\, \alpha(\omega,\omega') \:,\qquad
B^2(\omega, \omega') \;=\; -\omega'\, 4 a k M\, \alpha(\omega,\omega')\:,
\eeq
and~$\alpha$ as given by~(\ref{alphadef}).
The corresponding contribution to~(\ref{ins1}) can be estimated by
\beq \label{contri}
\int_{-\infty}^\infty d\omega\,|f h^a|(\omega)\, \|\phi^a_{\omega, n}\|_{L^\infty(\R)}\:
\int_{-\infty}^\infty \eta_K(u)\: |C(\omega, u)| \: du \: ,
\eeq
where
\begin{eqnarray*}
C(\omega, u) &:=& \int_{-\infty}^\infty g(\omega')\, h^b(\omega')\:(\omega-\omega')\:
\zeta_\delta(\omega-\omega') \\
&& \times \left( B^1(\omega, \omega') + B^2(\omega, \omega')\: \frac{r}{(r^2+a^2)^2} \right)\phi^b_{\omega', n'}(u)\, d\omega'\:.
\end{eqnarray*}
For~$l=1,2$ we introduce the functions
\beq \label{Fldef}
F^l_\omega(\omega') \;=\; g(\omega')\, h^b(\omega')\:(\omega-\omega')\:
\zeta_\delta(\omega-\omega') \: B^l(\omega, \omega')\:.
\eeq
We let~$\varepsilon = {\mbox{dist}}(0, {\mbox{supp}}\, g)$
and apply Lemma~\ref{lemmal1} to obtain the upper bound
\[ (\ref{contri}) \;\leq\;
\int_{-\infty}^\infty d\omega\,|f h^a|(\omega)\, \|\phi^a_{\omega, n}\|_{L^\infty(\R)}\;
c\,
\sum_{l=1}^2 \left( \|F^l_\omega\|_{H^{1,1}} \: \|F^l_\omega\|_{H^{2,1}} \right)^\frac{1}{2} \:, \]
independent of~$K$.
According to Lemma~\ref{lemmaangular}, the functions~$B^l$ and their derivatives are
polynomially bounded. The other functions in~(\ref{Fldef}) and their derivatives are
also polynomially bounded. Each time the function~$\zeta_\delta$ is differentiated,
the chain rule gives a factor~$1/\delta$. Finally, the factor~$\omega-\omega'$ is bounded
by~$\delta$. We conclude that
\[ \|F^l_\omega\|_{C^1} \;\leq\; {\mathcal{P}}(\omega, \omega')\:,\quad
\|F^l_\omega\|_{C^2} \;\leq\; \frac{{\mathcal{P}}(\omega, \omega')}{\delta} \]
with a suitable polynomial~${\mathcal{P}}$.
The $\omega'$-integration is taken over the interval~$[\omega-\delta, \omega+\delta]$,
giving rise to an additional factor of~$\delta$. We thus obtain
\[ (\ref{contri}) \;\leq\;
c\, \sqrt{\delta}\: \int_{-\infty}^\infty d\omega\,|f h^a|(\omega)\, \|\phi^a_{\omega, n}\|_{L^\infty(\R)}\: 
\sup_{\omega' \in [\omega-\delta, \omega+\delta]}
{\mathcal{P}}(\omega, \omega')\:. \]
Since~$f$ and~$\|\phi^a_{\omega, n}\|_{L^\infty(\R)}$
are bounded in~$\omega$ and~$h^a(\omega)$ has rapid decay, the integral is finite.
The lemma follows because~$\delta$ is arbitrary. 
\QED

Our next proposition also treats the case~$n=n'$
and~${\mbox{supp}}\, f \cap {\mbox{supp}}\, g \neq \emptyset$.
\begin{Prp} \label{prpfa}
Suppose that~$f \in C^{\infty}(\R)$ and~$g \in C^\infty_0(\R)$ with~$0 \not \in {\mbox{\rm{supp}}}\,f$
or~$0 \not \in {\mbox{\rm{supp}}}\,g$. Then
\[ \bra f(H)_n \Psi_0, g(H)_{n'} \Psi_0 \ket \;=\; \delta_{n n'}\:
\bra \Psi_0, ((f g)(H))_{n} \Psi_0 \ket\:. \]
\end{Prp}
{\Proof} In the case~$n \neq n'$ the statement follows immediately
from Proposition~\ref{lemma31}. Thus it remains to consider the case~$n=n'$.
If~$f=p$ is a polynomial, we sum over~$n$ and use that~$H$
is symmetric with respect to the energy inner product. This gives
\begin{eqnarray*}
\lefteqn{ \sum_{n \in \N} \bra p(H)_n \Psi_0, g(H)_{n'} \Psi_0 \ket \;=\;
\bra p(H) \Psi_0, g(H)_{n'} \Psi_0 \ket } \\
&=& \bra \Psi_0, p(H) g(H)_{n'} \Psi_0 \ket
\;=\; \bra \Psi_0, ((p g)(H))_{n'} \Psi_0 \ket \:.
\end{eqnarray*}
Since we already know that the summands with~$n \neq n'$ vanish, the
proposition holds for polynomials.

We choose a function~$\chi \in C^\infty_0(\R)$ which is identically equal
to one on the support of~$g$. By the Weierstra{\ss} approximation theorem,
there is sequence of polynomials~$p_j$ with~$p_j \rightarrow f$ 
in~$C^l({\mbox{supp}} \chi)$ (i.e.\ $p_j$ and its derivatives of order~$\leq l$
converge uniformly on~${\mbox{supp}} \,\chi$). Then, 
writing~$\chi p_j = p_j + (\chi-1) p_j$ and applying Proposition~\ref{lemma31}~{\bf{(i)}},
we obtain
\[ \bra ((\chi p_j)(H))_n \Psi_0, g(H)_n \Psi_0 \ket \;=\; \bra p_j(H)_n \Psi_0, g(H)_n \Psi_0 \ket \;=\;
\bra \Psi_0, ((p_j g)(H))_n \Psi_0 \ket \:. \]
The functions~$\chi p_j$ and~$p_j g$ converge in~$C^l(\R)$ to~$\chi f$ and~$f g$, respectively.
Corollary~\ref{cor1} allows us to pass to the limit to obtain
\[ \bra ((\chi f)(H))_n \Psi_0, g(H)_n \Psi_0 \ket \;=\;
\bra \Psi_0, ((f g)(H))_n \Psi_0 \ket\:. \]
Applying Proposition~\ref{lemma31}~{\bf{(i)}},
we conclude that the left side
equals $\bra f(H)_n \Psi_0, g(H)_n \Psi_0 \ket$. \\
\hspace*{1cm} \QED

\section{Energy Splitting and $L^2$-Bounds of the High Energy \\
Contribution} \label{secesplit}
\setcounter{equation}{0}
The next lemma allows us to restrict attention to a finite number of
angular momentum modes. We define
\begin{equation} \label{PsiNdef} \Psi_N(t,r, \vartheta) \;=\; 
\frac{1}{2 \pi} \sum_{n \leq N} \int_{-\infty}^\infty \frac{d\omega}{\omega
\Omega}\:e^{-i \omega t} \sum_{a,b=1}^2 t^{\omega n}_{ab}\:
\Psi_a^{\omega n}(r,\vartheta)\; \bra \Psi_b^{\omega n}, \Psi_0
\ket\:.
\end{equation}
\begin{Lemma} For any~$u_0 \in \R$ and $t>0$,
\[ \|\Psi(t)\|_{L^{2}((u_0, \infty) \times S^2)} \;\leq\; \limsup_{N \rightarrow \infty}
\|\Psi_N(t)\|_{L^{2}((u_0, \infty) \times S^2)} \:. \]
\end{Lemma}
{\Proof} Due to finite speed of propagation, the support of~$\Psi(t)$ lies in
a compact set~$K(t)$. Hence setting~$\Sigma = (u_0, \infty) \times S^2$,
\[ \|\Psi(t)\|_{L^2(\Sigma)} \;=\;
\|\Psi(t)\|_{L^2(K(t) \cap \Sigma)} \;=\; \lim_{N \rightarrow \infty}
\|\Psi_N(t)\|_{L^2(K(t) \cap \Sigma)}  , \]
where in the last step we used that~$\Psi_N$
converge in~$L^2_{\mbox{\scriptsize{loc}}}$ (see~\cite[Theorem~7.1]{FKSY2}).
We finally use that~$\|\Psi_N(t)\|_{L^2}$ increases
if the integration range is extended to~$\Sigma$.
\QED
This lemma applies just as well to weighted~$L^2$-norms,
and, as in the proof of Corollary~\ref{cor1}, it can be
extended to~$L^2$-estimates of the derivatives.
Hence this lemma also applies to the integral of the energy
density. Thus our task in proving Theorem~\ref{thm11} is to estimate
the energy of~$\Psi_N(t)$ in the region~$(u_0, \infty) \times S^2$
uniformly in~$N$ and~$t$.

We now perform an energy splitting decomposition similar to that in~\cite{FKSY3}.
For a given parameter~$J > 0$,
we introduce a cutoff function~$\chi_\low \in C^\infty_0([-2J,2J])$ which is identically
equal to one on the interval~$[-J,J]$. We set~$\chi_\high=1-\chi_\low$ and define
the low- and high-energy contributions by
\[ \Psi^\low_N \;=\; \chi_\low(H) \Psi_N \:,\spc \Psi^\high_N \;=\; \chi_\high(H) \Psi_N \:, \]
respectively. We now show that it suffices to estimate the low-energy contribution.
The energy of~$\Psi_N$ can be expressed by
\[ \bra \Psi_N, \Psi_N \ket \;=\; \bra \Psi^\high_N, \Psi^\high_N \ket
\:+\: 2 {\mbox{Re}}\, \bra \Psi^\high_N, \Psi^\low_N \ket \:+\:
\bra \Psi^\low_N, \Psi^\low_N \ket \:. \]
Using the functional calculus of Proposition~\ref{prpfa} gives for the mixed term
\[ \bra \Psi^\high_N, \Psi^\low_N \ket \;=\; \frac{1}{2 \pi}
\sum_{n \leq N} \int_{-\infty}^\infty \frac{d\omega}{\omega
\Omega}\:\chi_\high(\omega)\: \chi_\low(\omega) \sum_{a,b=1}^2 t^{\omega n}_{ab}\:
\bra \Psi_0, \Psi_a^{\omega n}\ket \; \bra \Psi_b^{\omega n}, \Psi_0 \ket\:. \]
Thus
\beq \label{psihighes}
\bra \Psi^\high_N, \Psi^\high_N \ket \;=\; \bra \Psi_N, \Psi_N \ket
\:-\: E^\low_N \:,
\eeq
where
\[ E^\low_N \;:=\; \frac{1}{2 \pi} \sum_{n \leq N} \int_{-\infty}^\infty \frac{d\omega}{\omega
\Omega}\:\chi_\low(\omega) \Big(\chi_\low(\omega) + 2 \chi_\high(\omega)
\Big) \sum_{a,b=1}^2 t^{\omega n}_{ab}\:
\bra \Psi_0, \Psi_a^{\omega n}\ket \; \bra \Psi_b^{\omega n}, \Psi_0 \ket\:. \]
Note that the integrand in~$E^\low_N$ is supported in the low-energy region~$[-2J, 2J]$.
If~$N \rightarrow \infty$, we know from~\cite{FKSY2} that~$\bra \Psi_N, \Psi_N \ket$
converges to the total energy of the initial data,
\[ \lim_{N \rightarrow \infty} \bra \Psi_N, \Psi_N \ket \;=\; \bra \Psi_0, \Psi_0 \ket\:. \]
Hence
\[ \limsup_{N \rightarrow \infty} \bra \Psi^\high_N, \Psi^\high_N \ket
\;=\; \bra \Psi_0, \Psi_0 \ket \:-\: \liminf_{N \rightarrow \infty} E^\low_N\:. \]
From~\cite[Theorem~1.1]{FKSY3} we conclude that there is a~$J>0$ such that
\beq \label{esin}
\limsup_{N \rightarrow \infty} \|\Phi^\high_N(t)\|^2_{L^2} \;\leq\; c
\left( \bra \Psi_0, \Psi_0 \ket \:-\: \liminf_{N \rightarrow \infty} E^\low_N \right) ,
\eeq
where~$c$ is independent of~$\Psi_0$.
Hence the~$L^2$-norm of the high-energy contribution, and thus also its
energy is under control, provided that we have an estimate for~$E^\low_N$.
The next two sections are concerned with estimating the low-energy
contributions~$\bra \Psi^\low_N, \Psi^\low_N \ket$ and~$E^\low_N$.

\section{Global ODE Estimates for Large~$\lambda$}
\setcounter{equation}{0}
We begin with a result which is a strengthened version of~\cite[Theorem~2.1]{FKSY3}.
To this end, we choose~$\delta \in (0, \frac{1}{8})$  and
fix~$u_0<0$ such that ${\mbox{supp}}\, \Psi_0 \subset \subset (u_0, \infty) \times S^2$.
As in~\cite{FKSY2}, we set~$\acute{\rho} = |\acute{\phi}|$, $\acute{y} =\acute{\phi}'/\acute{\phi}$,
and~$W = V + \Omega^2$. We define~$u_-, u_+ \ll 0$ and~$u_1, u_2 \gg 0$ implicitly by the relations
\[ V(u_-) \;=\; -\frac{\Omega^2}{2} \:,\qquad V(u_+) \;=\; \Omega^2\:,\qquad
V(u_1) \;=\; \delta^2 \omega^2\:,\qquad V(u_2) \;=\; -\delta^2 \omega^2 \:. \]
(Note that for sufficiently large~$\lambda$, the points~$u_-$, $u_+$, $u_1$ and~$u_2$ are
determined uniquely.)
\begin{Lemma} \label{lemma23}
For sufficiently large~$\lambda$,
there is a constant~$C>0$ such that for
all~$\omega \in (-2J, 2J) \setminus \{\omega_0, 0\}$ and
all~$n \geq n_0$ and $a=1,2$ the following inequalities hold:
\begin{eqnarray}
\frac{\sqrt{V}}{8} \:-\: C |\Omega| &\leq& {\mbox{\rm{Re}}}\, \acute{y}
\;\leq\; 2 \sqrt{V} \:+\: C |\Omega|
\qquad\;\;\: {\mbox{on $[2 u_0, u_1]$}} \label{nsum0}  \\
\frac{1}{|\Omega|} \sum_{b=1}^2 \left| t_{ab}^{\omega n}\:
\: \phi^b(u) \right| &\leq& \frac{C}{\acute{\rho}(u_0)} 
\quad\qquad\qquad\qquad\qquad\quad\! {\mbox{on $[u_0, u_1]$}} \label{nsum} \\
\frac{1}{|\Omega|} \sum_{b=1}^2 \left| t_{ab}^{\omega n}\:
\: \phi^b(u) \right| &\leq& \frac{C}{\acute{\rho}(u_1)}\: \frac{e^{3 \delta \sqrt{\lambda}}}
{\delta^2\, |\omega|} \qquad\qquad\qquad\,
\quad {\mbox{on $[u_1, \infty]$}} \label{nsum2} \:.
\end{eqnarray}
\end{Lemma}
{\Proof} Near infinity, the potential~$V$ has the asymptotic form
\beq \label{Vinf}
V(u) \;=\; -\omega^2 + \frac{\lambda}{u^2}\:+\: \lambda\, {\mathcal{O}}(u^{-3})\:,\qquad
V'(u) \;=\; - \frac{2 \lambda}{u^3}\:+\: \lambda\, {\mathcal{O}}(u^{-4})\:.
\eeq
A straightforward computation shows that
the assumptions of Lemma~4.12 in~\cite{FKSY2} are satisfied
on the interval~$[u_+, u_1]$ for sufficiently large~$\lambda$.
Hence~\cite[eq.~(4.40)]{FKSY2} continues to hold on~$[2 u_0, u_1]$,
and this gives the left inequality in~(\ref{nsum0}).
The right inequality in~(\ref{nsum0}) follows from~\cite[Lemma~4.2]{FKSY2}.

As in~\cite{FKSY3}, we can arrange by a phase change that~$\acute{\phi}(u_1)$ is real.
To prove~(\ref{nsum}), we first consider the case~$a=1$. Using the explicit formulas for~$t^{\omega n}_{1b}$
in the proof of Theorem~2.1 in~\cite{FKSY3}, we find that for any~$u \in [u_0, u_1]$
\[\frac{1}{|\Omega|} \sum_{b=1}^2  \left| t_{1b}^{\omega n}\:
\: \phi^b(u) \right| \;\leq\; \frac{1}{|\Omega|}
\left| 1 + \frac{\alpha}{\beta} \right| \acute{\rho}(u) \;\leq\;
\frac{4}{\acute{\rho}\, {\mbox{Re}}\, (\acute{y} - \grave{y})}
\;=\; \frac{4\, \sqrt{|{\mbox{Im}}\, \acute{y}}|}{\sqrt{|\Omega|}\, {\mbox{Re}}\, (\acute{y} -
\grave{y})} \:, \]
where we have applied~\cite[eq.~(18)]{FKSY3}.
Using the results in the proof of Theorem~2.1 in~\cite{FKSY3}, we
know that~${\mbox{Re}}\, (\acute{y} - \grave{y})$ is bounded away from
zero. Applying the the relation
\beq \label{wid}
\acute{\rho}^2 \;=\; \frac{\Omega}{{\mbox{Im}}\, \acute{y}} \:,
\eeq
we conclude that
\[ \frac{1}{|\Omega|} \sum_{b=1}^2 \left| t_{1b}^{\omega n}\:
\: \phi^b(u) \right| \;\leq\;
\frac{C}{\acute{\rho}(u)} \:. \]
Using that~$\acute{\rho}'/\acute{\rho} = {\mbox{Re}}\, \acute{y}$ and that~${\mbox{Re}}\, \acute{y} > 0$,
it follows that~$\acute{\rho}$ is monotone increasing on the interval~$[u_0, u_1]$.
This completes the proof of~(\ref{nsum}) on the interval~$[u_0, u_1]$ in the case~$a=1$.
In the case~$a=2$, the summand~$b=1$ can be estimated as above.
For the summand~$b=2$ we have
\begin{eqnarray*}
\left| \frac{1}{\Omega} \:t_{22}^{\omega n}\:
\: \phi_2(u) \right| \;\leq\; \frac{c}{|\Omega|}\:
|\phi_2(u)| \;\leq\; \frac{c \sqrt{|{\mbox{Im}} \,\acute{y}}|}{\sqrt{|\Omega|}}\, (u_1-u_0)\:,
\end{eqnarray*}
which is estimated just as above.

To prove~(\ref{nsum2}), we need
estimates for~$\grave{y}$. On the interval~$[u_2, \infty)$, we apply
the invariant region estimate of~\cite[Lemma~4.1]{FKSY2} with~$\alpha \equiv 0$
to conclude that on the interval~$[u_2, \infty)$,
\beq \label{inv1}
\delta \,|\omega| \;\leq\; |{\mbox{Im}}\, \grave{y}| \;\leq\; |\omega|\:,\qquad
|{\mbox{Re}}\, \grave{y}| \;\leq\; |\omega| \:.
\eeq
On the interval~$[u_1, u_2]$, we can apply~\cite[Lemma~4.10]{FKSY2}
setting~$\alpha=\sqrt{2 \delta^2 \omega^2}$. Using the asymptotic form of the
potential~$V$ near infinity~(\ref{Vinf}),
the points~$u_1$ and~$u_2$ can be estimated by
\beq \label{ues}
\frac{\sqrt{\lambda}}{2\, |\omega|} \;\leq\; u_1 < u_2 \;\leq\; \frac{2\, \sqrt{\lambda}}{|\omega|}\:.
\eeq
Furthermore, the mean value theorem yields that
\[ 2 \delta^2 \omega^2 = V(u_1) - V(u_2) \geq (u_2 - u_1)\, \inf_{[u_1, u_2]}|V'|\:, \]
and thus, using that~$|V'|$ is monotone decreasing,
\[ u_2 - u_1 \;\leq\; \frac{2 \delta^2 \omega^2}{|V'(u_2)|} \;\leq\;
\frac{2 \delta^2 \omega^2 u_2^3}{\lambda} \;\leq\; \frac{16 \, \sqrt{\lambda}\, \delta^2}{|\omega|}\:. \]
We conclude that
\[ 2 \alpha (u_2 - u_1) \;\leq\; 64 \: \sqrt{\lambda}\, \delta^3 \;\leq\; \sqrt{\lambda}\, \delta\:, \]
and thus on the interval~$[u_1, u_2]$,
\beq \label{inv2}
\frac{\delta}{C}\: e^{-\sqrt{\lambda}\, \delta}
\,|\omega| \;\leq\; |{\mbox{Im}}\, \grave{y}| \;\leq\; C\, e^{\sqrt{\lambda}\, \delta}\: |\omega|\:,\qquad
|{\mbox{Re}}\, \grave{y}| \;\leq\; C\, e^{\sqrt{\lambda}\, \delta}\:|\omega| \:,
\eeq
where~$C$ is independent of~$\omega$ and~$\lambda$.

We again consider the cases~$a=1$ and~$a=2$ separately. In the case~$a=2$,
we first express~$\phi_2$ in terms of~$\grave{\phi}$ and~$\overline{\grave{\phi}}$.
Making the ansatz
\[ \phi_2 \;=\; 2\, {\mbox{Re}}\, (\gamma \grave{\phi}) \:,\qquad
\phi_2' \;=\; 2\, {\mbox{Re}}\, (\gamma \grave{\phi}') \:, \]
the coefficient~$\gamma$ is computed to be
\[ \gamma \;=\; -\frac{\phi_2'(u_1)}{i \omega}\: \overline{\grave{\phi}(u_1)}\:. \]
Hence
\[ |\phi_2(u)| \;\leq\; \frac{2 |\phi_2'(u_1)|}{|\omega|}\: \grave{\rho}(u_1)\:\grave{\rho}(u) , \]
and using the identity~$\Omega = w(\phi_1, \phi_2)$ together with the
relations~$|\phi_1(u_1)|=\acute{\rho}(u_1)$ and $\phi_2(u_1)=0$, we obtain
\beq \label{phi2es}
|\phi_2(u)| \;\leq\; \frac{2 |\Omega|}{|\omega|}\:\frac{1}{\acute{\rho}(u_1)}\;
\grave{\rho}(u_1)\:\grave{\rho}(u) \:.
\eeq
Using the relation
\beq \label{65b}
\grave{\rho}^2 \;=\; \frac{|\omega|}{|{\mbox{Im}}\, \grave{y}|}
\eeq
as well as the estimates~(\ref{inv1}, \ref{inv2}),
we obtain~(\ref{nsum2}).

In the remaining case~$a=1$, we first express~$\phi_1$ in terms of~$\grave{\phi}$
and~$\overline{\grave{\phi}}$. The ansatz
\[ \phi_1 \;=\; 2\, {\mbox{Re}}\, (\gamma \grave{\phi}) \:,\qquad
\phi_2' \;=\; 2\, {\mbox{Re}}\, (\gamma \grave{\phi}') \]
yields
\[ \gamma \;=\; \frac{1}{i \omega} \left( \phi_1(u_1)\: \overline{\grave{\phi}}'(u_1) 
- \phi_1'(u_1)\: \overline{\grave{\phi}}(u_1) \right) . \]
Hence
\begin{eqnarray*}
\left| \frac{\phi_1(u)}{\phi_1(u_1)} \right|
&\leq& \frac{1}{|\omega|} \: |\grave{\phi}'(u_1)|\: \grave{\rho}(u)\:+\:
\frac{|\phi_1'(u_1)|}{|\phi_1(u_1)|}\: \frac{1}{|\omega|} \: \grave{\rho}(u_1)\: \grave{\rho}(u) \\
&\leq& \frac{1}{|\omega|}\: (|\grave{y}(u_1)| + |{\mbox{Re}}\,\acute{y}(u_1)|)\: \grave{\rho}(u_1)\: \grave{\rho}(u)\:, 
\end{eqnarray*}
where in the last step we used the relations~$\phi_1(u_1)=\acute{\rho}(u_1)$
and~${\mbox{Re}}\, \acute{\phi}' = {\mbox{Re}}(\phi_1'+i \phi_2') = \phi_1'$.
We now apply~(\ref{nsum0}) and~(\ref{65b}) as well as~(\ref{inv1}) and~(\ref{inv2}) to obtain
\[ \left| \frac{\phi_1(u)}{\phi_1(u_1)} \right| \;\leq\;
\frac{c}{|\omega|\, \delta^2}\: e^{3 \delta \sqrt{\lambda}}\:. \]
Combining this estimate with~(\ref{phi2es}) and again using~(\ref{65b})
and~(\ref{inv1}, \ref{inv2}), we get
\beq \label{quot}
\left| \frac{\acute{\rho}(u)}{\acute{\rho}(u_1)} \right| \;\leq\;
\frac{c}{|\omega|\, \delta^2}\: e^{3 \delta \sqrt{\lambda}}\:.
\eeq
We finally use the estimate
\[ \frac{1}{|\Omega|} \sum_{b=1}^2  \left| t_{1b}^{\omega n}\:
\: \phi^b(u) \right| \;\leq\; \frac{1}{|\Omega|}
\left| 1 + \frac{\alpha}{\beta} \right| \acute{\rho}(u) \;=\; 
\left\{ \frac{1}{|\Omega|} \left| 1 + \frac{\alpha}{\beta} \right| \acute{\rho}(u_1) \right\}
\left| \frac{\acute{\rho}(u)}{\acute{\rho}(u_1)} \right| . \]
The curly bracket term is estimated just as in the proof of~(\ref{nsum}), whereas
the last factor was estimated in~(\ref{quot}).
\QED
We next use the exponential increase of~$\acute{\rho}$ on the interval~$[u_0, u_1]$
to prove that the inequality in~(\ref{nsum}) holds all the way to infinity.
\begin{Lemma} \label{lemma231}
For sufficiently large~$\lambda$,
there is a constant~$C>0$ such that for
all~$\omega \in (-2J, 2J) \setminus \{\omega_0, 0\}$ and
all~$n \geq n_0$ the following inequality holds:
\[ \frac{1}{|\Omega|} \sum_{a=1}^2 \left| t_{ab}^{\omega n}\:
\phi_b(u) \right| \;\leq\; \frac{C}{\acute{\rho}(u_0)} 
\quad\qquad\qquad\;\;\: {\mbox{on $[u_0, \infty)$}} \:. \]
\end{Lemma}
{\Proof} In view of~(\ref{nsum2}) our task is to show that
\[ \frac{\acute{\rho}(u_1)}{\acute{\rho}(u_0)} \;\geq\;
\frac{e^{3 \delta \sqrt{\lambda}}}{\delta^2\, |\omega|}\:. \]
Equivalently, using that~$\acute{\rho}'/\acute{\rho} = {\mbox{Re}}\,
\acute{y}$, we need to show that the inequality
\beq \label{intes}
\int_{u_0}^{u_1} {\mbox{Re}}\, \acute{y} \;\geq\; 3 \delta \sqrt{\lambda}
- 2 \log \delta - \log |\omega|
\eeq
holds for sufficiently large~$\lambda$. On the interval~$[u_+, u_1]$
we can again use the invariant region estimate of~\cite[Lemma~4.12]{FKSY2}.
In particular, Lemma~4.2 in~\cite{FKSY2} holds on~$[u_+, u_1]$
for~$\alpha=7 \sqrt{V}/8$. Thus, using that the function~$\sigma^2 U$ in this lemma
is monotone increasing (cf~\cite[line after eq.~(4.24)]{FKSY2}) and
becomes large as~$\lambda \rightarrow \infty$, it is clear that
for sufficiently large~$\lambda$, the function $T$ appearing in~\cite[Lemma~4.2]{FKSY2}
will be smaller than one on the interval~$[u_0, u_1]$.
In other words, we are in the situation of~\cite[Fig.~2 (right)]{FKSY2} so that
\[ {\mbox{Re}}\, y \;\geq\; \alpha - \sqrt{U} \;\geq\; \frac{\sqrt{V}}{8}\:. \]
Using the asymptotic expansion~(\ref{Vinf}), we find
\[ \int_{u_0}^{u_1} {\mbox{Re}}\, y \;\geq\; \frac{\delta\, \sqrt{\lambda}}{16}
\left( \log u_1 - C \right)
\;\stackrel{(\ref{ues})}{\geq}\; \frac{\delta\, \sqrt{\lambda}}{16}\:
\log \left( \frac{\sqrt{\lambda}}{2\, |\omega|} - C \right) . \]
This implies that~(\ref{intes}) is valid for large~$\lambda$.
\QED

In the next lemma we estimate the error term of the
plane wave asymptotics near the event horizon.
\begin{Lemma} \label{lemma62}
There are constants~$c, \gamma>0$ such that for all~$\omega \in (-2J, 2J)$
and all sufficiently large~$\lambda$ the following inequality holds:
\[ \left| \acute{\phi}(u) - e^{i \Omega u} \right| \;\leq\; c\, e^{\frac{\gamma u}{2}} \:
\acute{\rho}(2 u_0) \spc {\mbox{for all~$u<2 u_0$}}\:. \]
\end{Lemma}
{\Proof} In the region~$u<u_-$, we use the invariant region estimate of Corollary~4.3 and
Corollary~4.9 in~\cite{FKSY2} to obtain
\beq \label{first}
\left| \acute{y} - i \Omega \right| \;\leq\; \frac{W}{\Omega}\:.
\eeq
Using the Taylor expansion of the proof of Lemma~4.8 in~\cite{FKSY2}, we have
\beq \label{second}
W(u) \;=\; (\lambda + c_0) \: e^{\gamma u} + {\mathcal{O}}(\lambda e^{2 \gamma u})\:.
\eeq
Hence for sufficiently large~$\lambda$, we know that for all~$u<u_-$,
\[ \frac{\Omega^2}{2} \;\geq\; W(u) \;\geq\; \frac{1}{2} \, (\lambda + c_0) \: e^{\gamma u} \]
and thus
\beq \label{exbound}
\frac{(\lambda + c_0) \: e^{\gamma u}}{\gamma \Omega}
\;\leq\; \frac{\Omega}{\gamma} \:.
\eeq
It follows that for all~$u<u_-$,
\begin{eqnarray}
|\acute{\phi}(u) - e^{i \Omega u}| &=& \left| \exp \left( \int_{-\infty}^u
\acute{y} - i \Omega \right) - 1 \right|
\;\stackrel{(\ref{first})}{\leq}\; \left| \exp \left( \int_{-\infty}^u \frac{W}{\Omega} \right) - 1 \right|
\nonumber \\
&\stackrel{(\ref{second})}{\leq}& \exp \left( \frac{2 (\lambda + c_0)}{\gamma \Omega}\: e^{\gamma u} \right) - 1
\;\leq\; \frac{C (\lambda + c_0)}{\gamma \Omega}\: e^{\gamma u}\:, \label{614b}
\end{eqnarray}
where in the last step we used that the argument of the exponential is bounded
in view of~(\ref{exbound}). Substituting in the square root of~(\ref{exbound}), we conclude that there
is a constant~$c$ such that for all sufficiently large~$\lambda$,
\beq \label{um}
|\acute{\phi}(u) - e^{i \Omega u}| \;\leq\; c\, \sqrt{\lambda}\, e^{\frac{\gamma u}{2}}
\qquad {\mbox{for all~$u \leq u_-$}}\:.
\eeq

On the interval~$[u_-, u_+]$ we can apply the method of~\cite[Corollary~4.11]{FKSY2}. More
precisely, we use the estimate
\[ \left| \int_{u_-}^{u_+} \acute{y} - i \Omega \right| \;\leq\;
c\, |\Omega|\, (u_+-u_-) \:. \]
Since~$(u_+-u_-)$ is bounded according to~\cite[eq.~(4.29)]{FKSY2} and~$\Omega$ is bounded
in the low-energy region, we see that
(possibly after increasing~$c$) the inequality~(\ref{um}) is also valid on the interval~$[u_-, u_+]$.

For a sufficiently large constant~$D>0$, we introduce~$\hat{u} \ll 0$ by the condition
\beq \label{hudef}
V(\hat{u}) \;=\; D
\eeq
(as~$V(u_0)$ tends to infinity as~$\lambda \rightarrow \infty$, by increasing~$\lambda$
we can always arrange that this equation has a unique solution on~$(-\infty, 2 u_0]$).
Thus we have the relations
\[ u_- \;<\; u_+ \;<\; \hat{u} \;<\; 2 u_0 \qquad {\mbox{and}} \qquad
{\mbox{supp}}\, \Psi_0 \subset [u_0, \infty) \times S^2 \:. \] 
On the interval~$[u_+, \hat{u}]$,
\begin{eqnarray}
|\acute{\phi}(u) - e^{i \Omega u}| &=& \left| e^{\int_{-\infty}^u (\acute{y} - i \Omega)} - 1 \right|
\;\leq\; \left| e^{\int_{-\infty}^u (\acute{y} - i \Omega)} - e^{\int_{-\infty}^{u_+} (\acute{y} - i \Omega)}
\right| + \left| e^{\int_{-\infty}^{u_+} (\acute{y} - i \Omega)} - 1 \right| \nonumber \\
&=& \left| e^{\int_{-\infty}^{u_+} (\acute{y} - i \Omega)} \right|
\left| e^{\int_{u_+}^u (\acute{y} - i \Omega)} - 1 \right| +
\left| e^{\int_{-\infty}^{u_+} (\acute{y} - i \Omega)} - 1 \right| \nonumber \\
&=& \left| \acute{\phi}(u_+)\right|
\left| e^{\int_{u_+}^u (\acute{y} - i \Omega)} - 1 \right| +
\left| \acute{\phi}(u_+) - e^{i \Omega u_+} \right| . \label{rest}
\end{eqnarray}
Using~(\ref{614b}, \ref{exbound}) and~(\ref{um}) at~$u_+$,
we conclude that on the interval~$[u_+, \hat{u}]$,
\beq \label{611a}
|\acute{\phi}(u) - e^{i \Omega u}| \;\leq\;
c \left| e^{\int_{u_+}^u (\acute{y} - i \Omega)} - 1 \right| +
c\, \sqrt{\lambda}\, e^{\frac{\gamma u_+}{2}}\:.
\eeq
To estimate the integral term, we apply the invariant region estimates of
Lemma~4.2 and Lemma~4.12 of~\cite{FKSY2}, which
are\footnote{We note that there is a typo
in~\cite[eq.~(4.37)]{FKSY2}, where the term~$T_0/2$ should be replaced by~$T_0 |\Omega|/2$.}
\begin{eqnarray}
\frac{\sqrt{V}}{8} - \frac{T_0\, |\Omega|}{2} &\leq& {\mbox{Re}}\, \acute{y} \;\leq\; c \,\sqrt{V}
\label{in1} \\
{\mbox{Im}}\, \acute{y}(u) &\leq& |\Omega| \, \exp \left( -\frac{7}{4} \int_{u_+}^u \sqrt{V} \right) ,  \label{in2}
\end{eqnarray}
valid on the interval~$[u_+, 2 u_0]$.
Using the right inequality in~(\ref{in1}) and~(\ref{in2}) together with~(\ref{second}), we obtain
\begin{eqnarray*}
\left| \int_{u_+}^u \acute{y} - i \Omega \right| &\leq&
\int_{u_+}^u \left( |\Omega| + c \sqrt{V} \right) \;\leq\;
c \int_{u_+}^u \left( |\Omega| + \sqrt{\lambda} \, e^{\frac{\gamma u}{2}}\right) \\
&=& c |\Omega|\, (u-u_+)  + 2c\, \frac{\sqrt{\lambda}}{\gamma}
\left( e^{\frac{\gamma u}{2}} - e^{\frac{\gamma u_+}{2}} \right) \\
&\leq& C\, |\Omega| + 2c\, \frac{\sqrt{\lambda}}{\gamma}\: e^{\frac{\gamma u}{2}} \:,
\end{eqnarray*}
where in the last line we have dropped negative terms and used that~$\hat{u}$ 
as defined by~(\ref{hudef}) is
uniformly bounded from above. Applying the estimate
\[ e^{\frac{\gamma u}{2}} \;\geq\; e^{\frac{\gamma u_+}{2}} \;
\stackrel{(\ref{second})}{\geq}\; \frac{|\Omega|}{2 \sqrt{\lambda}}\:, \]
we conclude that
\beq \label{intbound}
\left| \int_{u_+}^u \acute{y} - i \Omega \right| \;\leq\;
C\, \sqrt{\lambda}\: e^{\frac{\gamma u}{2}}\:.
\eeq
By definition of~$\hat{u}$, (\ref{hudef}), we know that~$\sqrt{\lambda} e^{\frac{\gamma u}{2}} \sim \sqrt{D}$,
and thus~(\ref{intbound}) is uniformly bounded. Hence we can apply~(\ref{intbound}) in~(\ref{611a})
to obtain together with~(\ref{um}) the inequality
\begin{eqnarray}
|\acute{\phi}(u) - e^{i \Omega u}| &\leq& c\, \sqrt{\lambda}\: e^{\frac{\gamma u}{2}} \nonumber \\
&=& c\, \left( \frac{\sqrt{\lambda}}{\acute{\rho}(2 u_0)}
\right) e^{\frac{\gamma u}{2}}\: \acute{\rho}(2 u_0) \qquad {\mbox{for all~$u < \hat{u}$}}.
\label{es1}
\end{eqnarray}

On the remaining interval~$[\hat{u}, 2 u_0]$, the factor~$e^{\frac{\gamma u}{2}}$ in the
statement of the lemma is bounded from below and can be discarded. We use the estimate
\beq \label{es2}
|\acute{\phi}(u) - e^{i \Omega u}| \;\leq\; \acute{\rho}(u) + 1 \;\leq\;
\left( \frac{\acute{\rho}(u)}{\acute{\rho}(2 u_0)} + \frac{1}{\acute{\rho}(2 u_0)} \right)
\acute{\rho}(2 u_0) \qquad {\mbox{for all~$u \in [\hat{u}, 2 u_0]$}}. 
\eeq

To bound the brackets in~(\ref{es1}) and~(\ref{es2}), we use the estimates
\begin{eqnarray*}
\acute{\rho}(2 u_0)^2 &=& \frac{\Omega}{{\mbox{Im}}\, y(2 u_0)} \;\stackrel{(\ref{in2})}{\geq}\;
\frac{1}{c} \, \exp \left(\frac{7}{4} \int_{u_+}^{2 u_0} \sqrt{V} \right) \;\geq\; \frac{1}{c}\,
e^{\varepsilon \sqrt{\lambda}} \\
\log \frac{\acute{\rho}(2 u_0)}{\acute{\rho}(u)} &=& 
\int_{u}^{2 u_0} {\mbox{Re}}\, \acute{y} \;\stackrel{(\ref{in1})}{\geq}\;
\int_u^{2 u_0} \left( \frac{\sqrt{V}}{8} - \frac{T_0\, |\Omega|}{2} \right)
\;\geq\; \int_u^{2 u_0} \frac{\sqrt{V}}{16} \;>\;0 \:,
\end{eqnarray*}
valid for any~$u \in [\hat{u}, 2 u_0]$, where in the last line we chose~$D$ sufficiently large.
\QED

\section{$L^2$-Bounds of the Low-Energy Contribution Near the \\
Event Horizon}
\setcounter{equation}{0} \label{sec7}
To estimate the integral of the energy density of~$\Psi^\low_N(t)$ near infinity,
we first note that the local decay property~\cite{FKSY2} implies that the
integral of the energy density over any compact set is bounded uniformly in time.
Furthermore, from~\cite[Theorem~2.1]{FKSY3} we know that the energy of~$\Psi^\low_N$
as well as~$E_N^\low$ are finite, for example
\begin{eqnarray*}
|\bra \Psi^\low_N, \Psi^\low_N \ket| &\leq&
\int_{-\infty}^\infty d\omega
\:\chi_\low(\omega)^2
\;\sum_{n=1}^\infty  \left| \frac{1}{\Omega} \sum_{a,b=1}^2 t^{\omega n}_{ab}\:
\, \frac{\bra \Psi_0, \Psi_a^{\omega n}\ket}{\omega} \; \bra \Psi_b^{\omega n}, \Psi_0 \ket \right| \\
&\leq& 4 J \sup_{\omega \in [-2J, 2J]}
\left| \frac{1}{\Omega} \sum_{a,b=1}^2 t^{\omega n}_{ab}\:
\, \frac{\bra \Psi_0, \Psi_a^{\omega n}\ket}{\omega} \; \bra \Psi_b^{\omega n}, \Psi_0 \ket \right|
\;\leq\; C \,.
\end{eqnarray*}
Using conservation of energy, it thus suffices to estimate the integral of the energy
density near the event horizon. The method of the proof of Corollary~\ref{cor1}
shows that a time-independent~$L^2$-bound immediately gives rise to a corresponding~$H^{s,2}$-bound,
which dominates the integral of the energy density from above.
Hence it suffices to estimate the~$L^2$-norm near the event horizon, i.e.\ for any given~$u_0 \ll 0$
we must prove the inequality
\[ \|\Psi^\low_N(t)\|_{L^2((-\infty, 2 u_0) \times S^2)} \;\leq\; C \]
with~$C$ independent of~$N$ and~$t$.

Now using the notation in~(\ref{fndef})
with~$f(\omega) = e^{-i \omega t}\, \chi_\low(\omega)$, we apply the Schwarz
inequality to obtain
\begin{eqnarray*}
\|\Psi^\low_N(t)\|^2_{L^2((-\infty, 2 u_0) \times S^2)} &=& \sum_{n,n'=1}^N
\langle (e^{-i t H} \chi_\low(H))_n \Psi_0, (e^{-i t H} \chi_\low(H))_{n'} \Psi_0
\rangle_{L^2} \\
&\leq& \sum_{n,n'=1}^N
\left\| (e^{-i t H} \chi_\low(H))_n \Psi_0 \right\|_{L^2}\:
\left\| (e^{-i t H} \chi_\low(H))_{n'} \Psi_0 \right\|_{L^2} \\
&=& \left( \sum_{n=1}^N \| (e^{-i t H} \chi_\low(H))_n \Psi_0\|_{L^2} \right)^2\:.
\end{eqnarray*}
Furthermore, using the single mode estimates of Lemma~\ref{lemma2}, we may
disregard a finite number of angular momentum modes. 
Hence it remains to show that for any given~$n_0$,
\beq \label{task}
\sum_{n=n_0}^N
\left\| (e^{-i t H} \chi_\low(H))_n \Psi_0 \right\|_{L^2((-\infty,2 u_0) \times S^2)}
\;\leq\; C(\Psi_0, u_0, n_0)
\eeq
with the constant~$C$ independent of~$N$ and~$t$.
From now on, we will use the abbreviated notation
\beq \label{Psint}
\Psi^\low_n(t) \;=\; (e^{-i t H} \chi_\low(H))_n \Psi_0\:.
\eeq
This function has the following integral representation,
\beq \label{psin}
\Psi^\low_n(t,u,\vartheta) \;=\; \frac{1}{2 \pi} \int_{-2J}^{2J} \frac{1}{\omega \Omega}
\:e^{-i \omega t}\: \chi_\low(\omega)\:
t_{ab}^{\omega n} \:\Psi_a^{\omega n}(u,\vartheta)\: \bra \Psi_b^{\omega n}, \Psi_0 \ket\: d\omega \:.
\eeq

We next consider the behavior of~(\ref{psin}) near the event horizon.
Using~(\ref{aprep}), the plane wave asymptotics of the corresponding function~$\phi_n$
near the event horizon is given by
\[ \phi^{\mbox{\scriptsize{asy}}}_n(t,u,\vartheta) \;:=\; \int_{-2J}^{2J}
\left( \cos(\Omega u)\: h_1(\omega) + \sin(\Omega u)\: h_2(\omega) \right)\: d\omega \:, \]
where
\[ h_a(\omega) \;=\; \frac{1}{2 \pi\, \omega \Omega}
\:e^{-i \omega t}\: \chi_\low(\omega)\:
t_{ab}^{\omega n} \: \bra \Psi_b^{\omega n}, \Psi_0 \ket\:. \]
The~$L^2$-norm of~$\Phi^{\mbox{\scriptsize{asy}}}_n$ can be computed from Plancherel's theorem
to be
\begin{eqnarray*}
\|\Phi^{\mbox{\scriptsize{asy}}}_n\|^2_{L^2((-\infty, 2 u_0) \times S^2)} &\leq&
c(u_0)\: \|\phi^{\mbox{\scriptsize{asy}}}_n\|^2_{L^2(\R)} \;=\;
\pi\,c(u_0) \int_{-2J}^{2J}\left( |h_1(\omega)|^2 + |h_2(\omega)|^2 \right) d\omega \\
&\leq& c \int_{-2J}^{2J}  \frac{\chi_\low(\omega)^2}{(\omega \Omega)^2} \:
\bra \Psi_0, \Psi_a^{\omega n} \ket\: t_{ab}^{\omega n}
\,t_{bc}^{\omega n} \: \bra \Psi_c^{\omega n}, \Psi_0 \ket\, d\omega\:. 
\end{eqnarray*}
Hence
\begin{eqnarray*}
\|\Phi^{\mbox{\scriptsize{asy}}}_n\|^2_{L^2((-\infty, 2 u_0) \times S^2)} &\leq&
c \int_{-2J}^{2J}  \chi_\low(\omega)^2 \sum_b \left| \frac{1}{\omega \Omega} \:
t_{bc}^{\omega n} \: \bra \Psi_c^{\omega n}, \Psi_0 \ket \right|^2 d\omega \\
&\leq& c\: 4 J \: \frac{C^2}{\acute{\rho}(2 u_0)^2} \;\leq\;
c\: 4 J \: C^4\: e^{-2 \varepsilon n} \:,
\end{eqnarray*}
where in the last line we applied Lemma~\ref{lemma231}. We conclude that
\[ \sum_{n \geq n_0}^N \|\Phi^{\mbox{\scriptsize{asy}}}_n\|_{L^2((-\infty, 2 u_0) \times S^2)}
\;\leq\; C \]
with~$C$ independent of~$t$ and~$N$. Hence the plane wave asymptotics of~$\Psi_n(t)$ is under control,
and it remains to estimate the contribution of the error term~$E_\omega(u)$ in~(\ref{aprep})
to~$\Psi_n(t)$. For this we introduce the function
\[ \phi^{\mbox{\scriptsize{err}}}_n(t,u,\vartheta) \;=\; \int_{-2J}^{2J}
\left( {\mbox{Re}}\, (E_\omega)\: h_1(\omega) + {\mbox{Im}}\, (E_\omega)\: h_2(\omega) \right)\: d\omega \:. \]

We can now complete the proof of Theorem~\ref{thm11}.
Subtracting the plane wave asymptotics from~(\ref{psin}), we can apply Lemma~\ref{lemma231}
to obtain
\begin{eqnarray*}
\left|\phi^{\mbox{\scriptsize{err}}}_n(t,u,\vartheta) \right| &\leq& C \int_{-2J}^{2J}
\chi_\low(\omega)\:
|\acute{\phi}^{\omega n}(u) - e^{i \Omega u}| \:\frac{C}{\acute{\rho}(u_0)}\:
d\omega \\
&\leq& 4 J C \sup_{\omega \in (-2J, 2J)} |\acute{\phi}^{\omega n}(u) - e^{i \Omega u}|\;
\:\frac{C}{\acute{\rho}(u_0)}\:.
\end{eqnarray*}
Applying Lemma~\ref{lemma62}, we obtain for all~$n>n_0$ and $u<2 u_0$,
\[ \left|\phi^{\mbox{\scriptsize{err}}}_n(t,u,\vartheta) \right| \;\leq\;
4 J C^2 \:c\, e^{\frac{\gamma u}{2}} \:
\frac{\acute{\rho}(2 u_0)}{\acute{\rho}(u_0)}\:. \]
We now use that
\[ \frac{\acute{\rho}(u_0)}{\acute{\rho}(2 u_0)} \;=\;
\exp \left( \int_{2 u_0}^{u_0} {\mbox{Re}}\, \acute{y}\, du \right) \;\geq\;
c\; e^{\varepsilon \sqrt{\lambda_n(\omega)}} \:, \]
where in the last step we applied~(\ref{nsum0}). Using Weyl's asymptotics~$\lambda_n \sim n^2$
(see~\cite{FS}), we conclude that
\[ \left|\phi^{\mbox{\scriptsize{err}}}_n(t,u,\vartheta) \right| \;\leq\;
c\, e^{\frac{\gamma u}{2}} \:e^{- \varepsilon n} \:; \]
this gives
\[ \sum_{n \geq n_0}^N \|\Psi^{\mbox{\scriptsize{err}}}_n(t) \|_{L^2((-\infty, 2 u_0) \times S^2)}
\;\leq\; c \sum_{n \geq n_0}^N \|\phi^{\mbox{\scriptsize{err}}}_n(t) \|_{L^2((-\infty, 2 u_0))}
\;\leq\; \frac{2 c^2}{\gamma}\; e^{\frac{\gamma u_0}{2}} \sum_{n \geq n_0}^N e^{- \varepsilon n}
\;\leq\; C \]
with~$C$ independent of~$t$ and~$N$. This gives~(\ref{task}) and thus completes the proof of
Theorem~\ref{thm11}. \\

\section{Application to Wave-Packet Initial Data}
\setcounter{equation}{0}
In this section we consider
wave-packet initial data of the form
\begin{eqnarray} \label{wavepack}
\Psi_0^k \;=\; \left( \!\begin{array}{cc} \Phi_{0}\\
i \partial_{t}\Phi_{0}\end{array}\! \right)
\;=\; \Theta_{{\tilde n},{\tilde
\omega}}(\vartheta)\: \frac{\eta_{L}(u)}{\sqrt{r^{2}+a^{2}}}
\left[ \cin\,e^{-i{\tilde \omega}u}  \left( \!\begin{array}{cc}
1 \\ {\tilde \omega} \end{array}\! \right)
+ \cout\,e^{i{\tilde \omega}u}  \left( \!\begin{array}{cc}
1 \\ -{\tilde \omega} \end{array}\! \right) \right]
\end{eqnarray}
for fixed parameters~$k, \tilde{\omega}, \tilde{n}$ and~$\cin, \cout$.
Here~$L>0$ and
\[ \eta_{L}(u)\;=\;\frac{1}{\sqrt L}\,\eta \!\left(\frac{u-L^2}{L} \right) , \]
where $\eta \in C^\infty_0([-1,1])$ is a smooth cut-off function.
As~$L$ increases, the wave packet spreads out and at the same time moves
towards infinity. In~\cite{FKSY4} we need the following uniform in~$L$ result.
\begin{Thm} \label{thm2}
Consider the Cauchy problem~(\ref{cauchy}) for wave packet initial data~(\ref{wavepack}).
For every~$\varepsilon>0$ there is an~$n_0 \in N$ such that the
angular momentum modes~$\Phi_n$ of the corresponding solution satisfy
for all~$R>r_1$ and~$L$ the bound
\[ \limsup_{t \rightarrow \infty}
\int_{[R, \infty) \times S^2} {\mathcal{E}}\left(
\sum_{n=n_0}^\infty \Phi_n(t,x) \right)\, dr\, d\varphi\, d\cos \vartheta
\;\leq\; \varepsilon \:. \]
\end{Thm}
In preparation for the proof of this theorem, we need two lemmas.
\begin{Lemma} \label{lemma82}
Consider the Cauchy problem~(\ref{cauchy}) for wave packet initial data~(\ref{wavepack}).
For every~$\varepsilon>0$ there is an~$n_0 \in N$ such that the
angular momentum modes~$\Phi_n$ of the corresponding solution satisfy
for all~$L$ the bound
\[ \sum_{n=n_0}^\infty \left\|
 \Phi^\low_n(t) \right\|_{L^2(-\infty, 2 u_0) \times S^2)} \;\leq\; \varepsilon \:. \]
\end{Lemma}
{\Proof} Using that the amplitude of the wave packet~$\Psi_0$ scales like~$1/\sqrt{L}$, whereas
the size of its support scales like~$L$, the estimate of Lemma~\ref{lemma231} yields
\beq \label{Lest}
\frac{1}{|\Omega|} \sum_{b=1}^2 \left| t^{\omega n}_{ab} \, \bra \Psi^{\omega n}_b, \Psi_0^k \ket \right|
\;\leq\; \frac{C}{\acute{\rho}(u_0)}\: \sqrt{L}\:.
\eeq
Keeping track of the~$L$-dependence in Section~\ref{sec7}, we obtain the bound
\[ \|\Phi^\low_n \|_{L^2((-\infty, 2 u_0) \times S^2)} \;\leq\;
c\, e^{-\varepsilon \sqrt{\lambda_n}}\: \sqrt{L}\:. \]
The modes with~$\sqrt{\lambda_n} > (\log L) / \varepsilon$ can be controlled by the estimate
\[ e^{-\varepsilon \sqrt{\lambda_n}}\: \sqrt{L} \;=\; e^{-\frac{\varepsilon n}{2}}\: (e^{-\frac{\varepsilon \sqrt{\lambda_n}}{2}}\sqrt{L})
\;\leq\; e^{-\frac{\varepsilon n}{2}} \:. \]
Hence it remains to consider the modes with~$n \geq n_0$ and~$\lambda_n < (\log L)^2/\varepsilon^2$.
Furthermore, splitting up the $\omega$-integrals in Section~\ref{sec7} as
\[ \int_{-2J}^{2J} \ldots \,d\omega \;=\; \int_{[-2J,2J] \setminus [-L^{-1}, L^{-1}]} \ldots \,d\omega
\:+\: \int_{-1/L}^{1/L} \ldots \,d\omega\:, \]
the last integral can be estimated by $2/L$ times the supremum of the integrand.
The resulting factors~$1/L$ compensate the factors~$\sqrt{L}$ in~(\ref{Lest}). Hence we may
assume that~$|\omega| > 1/L$. Finally, noting that the support of~$\eta_L$
lies in the interval~$[L^2-L, L^2+L]$, we conclude that we can make the following assumptions
for large~$L$,
\beq \label{assum}
u \in \left[\frac{L^2}{2}, 2 L^2 \right]\:,\spc |\omega| \;\geq\; \frac{1}{L} \:,\spc
\lambda_n \;<\; \frac{\log^2 L}{\varepsilon^2} \:.
\eeq

Next we express~$t^{\omega n}_{ab} \phi_b$ in terms of the
fundamental solution~$\grave{\phi}$. According to~(\ref{gdef}, \ref{tabrel}),
\[ t_{ab} \phi_a(u)\, \phi_b(v) \;=\; -2 \Omega \; {\mbox{Im}} \left( \frac{\acute{\phi}(u)\, \grave{\phi}(v)}
{w(\acute{\phi}, \grave{\phi})} \right) \;=\;
-2 \Omega \left( \phi_1(u)\: {\mbox{Im}}\, \frac{\grave{\phi}(v)}
{w(\acute{\phi}, \grave{\phi})} + \phi_2(u) \: {\mbox{Re}}\, \frac{\grave{\phi}(v)}
{w(\acute{\phi}, \grave{\phi})} \right) . \]
Hence
\beq \label{tabrel2}
t_{1b} \, \phi_b \;=\; -2 \Omega\, {\mbox{Im}}\, \frac{\grave{\phi}}
{w(\acute{\phi}, \grave{\phi})} \:,\qquad
t_{2b} \, \phi_b \;=\; -2 \Omega\, {\mbox{Re}}\, \frac{\grave{\phi}}
{w(\acute{\phi}, \grave{\phi})} \:,
\eeq
and thus for~$a=1,2$,
\[ \left| t_{ab} \, \phi_b(v) \right| \;\leq\; \frac{2 \,|\Omega\, \grave{\phi}(v)|}
{|w(\acute{\phi}, \grave{\phi})|}\:. \]

In order to estimate~$\grave{\phi}$, we use the decomposition
\[ \grave{\phi} \;=\; e^{-i \omega u} + E(u) \:. \]
The error term can be controlled using the estimates for the Jost functions of~\cite[Lemma~3.3]{FKSY2}
\[ |E(u)| \;\leq\; \frac{\lambda_n}{|\omega|\, u} \;\stackrel{(\ref{assum})}{\leq}\; 
\frac{2 \log^2 L}{L \varepsilon^2} \:. \]
The extra decay in~$L$ can be used to compensate the factor~$\sqrt{L}$ in~(\ref{Lest}), and thus
the contribution of the error terms to~(\ref{Lest}) tends to zero as~$L \rightarrow \infty$.
Hence it remains to consider the plane-wave contribution
\beq \label{enexp}
\frac{1}{|\Omega|} \sum_{b=1}^2 \left| t^{\omega n}_{ab} \, \bra 
(r^2+a^2)^{-\frac{1}{2}} \left( \!\begin{array}{c} 1 \\ \omega \end{array} \!\right) e^{-i \omega u}\:
\Theta_{n,\omega}, \Psi_0^k \ket \right| \:.
\eeq
Writing the energy inner product in~(\ref{enexp}) using~(\ref{eipr}), 
we obtain expressions, one of which is of the form
\beq \label{planein}
\langle \Theta_{n,\omega}, \Theta_{\tilde{n}, \tilde{\omega}} \rangle_{L^2(S^2)}
\int_{-\infty}^\infty e^{i (\omega-\tilde{\omega}) u}\: \tilde{\eta}_L(u)\, du \:,
\eeq
where the function~$\tilde{\eta}_L$ is composed of~$\eta_L$ and its first two derivatives (and powers
of~$\tilde{\omega}$ and~$L^{-1}$). We integrate by parts and apply Lemma~\ref{lemmaangular} to obtain
in the case~$n \neq \tilde{n}$
\[ (\ref{planein}) \;=\; i \alpha(\omega, \tilde{\omega}) 
\int_{-\infty}^\infty e^{-i (\omega-\tilde{\omega}) u}\: \tilde{\eta}_L'(u)\, du\:. \]
Now the derivative~$\tilde{\eta}_L'$ is of the order~${\mathcal{O}}(L^{-1})$,
compensating the factor~$\sqrt{L}$ in~(\ref{Lest}).
Finally, we also get expressions of the form~(\ref{planein}), but with~$\tilde{\omega}$ replaced
by~$-\tilde{\omega}$. In this case, the inner product yields a factor~$(\omega+\tilde{\omega})$,
which allows us to again apply the above integration by parts method.
\QED

\begin{Lemma} \label{lemma83}
Consider the Cauchy problem~(\ref{cauchy}) for initial data~(\ref{wavepack}).
For every~$\varepsilon>0$ there is an~$n_0 \in N$ such that for any~$f \in C^\infty(\R)$
and sufficiently large~$L$,
\[ \sum_{n=n_0}^\infty |\bra \Psi_0, f(H)_n \Psi_0 \ket| \;\leq\; \varepsilon\, \sup |f|\:. \]
\end{Lemma}
{\Proof} We rewrite the wave propagator purely in terms of the fundamental solutions~$\grave{\phi}$.
A straightforward computation using~(\ref{trans}) and the identity (see~\cite[eq.~(6.2)]{FKSY2})
\[ |\alpha|^2 - |\beta|^2 \;=\; - \frac{\omega}{\Omega} \:, \]
together with~(\ref{gdef}, \ref{tabrel}) yields that
\[ \frac{1}{\omega \Omega} \: t^{\omega n}_{ab} \:\Phi_a(u) \,\Phi_b(v) \;=\;
\langle \grave{\bf{\Phi}}(u), {{\bf{U}}} \grave{\bf{\Phi}}(v) \rangle_{\C^2} \:, \]
where we use the matrix notation
\[ \grave{\bf{\Phi}} \;=\; \left( \! \begin{array}{c}
\grave{\Phi} \\ \overline{\grave{\Phi}} \end{array} \! \right)\:,\qquad
{\bf{U}} \;=\; \frac{1}{2 \omega^2} \left( \!\begin{array}{cc}
1 & -\overline{\alpha}/\beta \\
-\alpha/\overline{\beta} & 1 \end{array}\! \right) . \]
We decompose~$U$ into the difference of two positive operators,
\[ {\bf{U}} \;=\; {\bf{U}}_+ - {\bf{U}}_- \:, \]
where
\[ {\bf{U}}_- \;=\; \frac{1}{4 \omega^2} \left(\left|\frac{\alpha}{\beta} \right|-1 \right)
\left( \!\begin{array}{cc}
1 & \displaystyle \frac{\overline{\alpha \beta}}{|\alpha \beta|} \\
\displaystyle \frac{\alpha \beta}{|\alpha \beta|}
& 1 \end{array}\! \right) \qquad {\mbox{if $\omega \Omega < 0$}} \]
and~${\bf{U}}_-=0$ otherwise. Note that~${\bf{U}}_-$ vanishes outside the
interval~$[-2J, 2J]$ (with~$J$ chosen as in Section~\ref{secesplit}).

The norm of~${\bf{U}}_-$ can be estimated using the
first inequality in~\cite[Lemma~6.1]{FKSY2} as
\[ \|U_-\| \;\leq\; \frac{1}{2 \omega^2} \left| \left|\frac{\alpha}{\beta} \right| - 1 \right|
\;\leq\; \frac{|\Omega|}{|\omega|\, |w(\acute{\phi}, \grave{\phi})|^2}\:, \]
and thus from~(\ref{tabrel2})
\begin{eqnarray*}
\lefteqn{ |\langle \grave{\bf{\Phi}}^{\omega n}(u), {{\bf{U}}}^{\omega n}_-
\grave{\bf{\Phi}}^{\omega n}(v) \rangle_{C^2}| \;\leq\;
2 \| {{\bf{U}}}^{\omega n}_-\|\, |\grave{\Phi}^{\omega n}(u)|\, |\grave{\Phi}^{\omega n}(v)| } \\
&\leq& \frac{2 |\Omega|}{|\omega|}\: \left| \frac{\grave{\Phi}^{\omega n}(u)}{w(\acute{\phi}, \grave{\phi})}
\right| \left| \frac{\grave{\Phi}^{\omega n}(v)}{w(\acute{\phi}, \grave{\phi})}
\right| \;\leq\;
\frac{2 |\Omega|}{|\omega|} \left| \sum_a \frac{t_{ab}}{\Omega}\: \Phi_b(u) \right|
\left| \sum_a \frac{t_{ab}}{\Omega}\: \Phi_b(v) \right| .
\end{eqnarray*}
The resulting expression can now be estimated
as in the proof of Lemma~\ref{lemma82} to conclude that
\beq \label{Umies}
\sum_{n \geq n_0} \left| \int_{-2J}^{2J} f(\omega) \, \langle \grave{\bf{\Psi}}_0^{\omega n},
{{\bf{U}}}^{\omega n}_- \grave{\bf{\Psi}}_0^{\omega n} \rangle_{\C^2} \right|
\;\leq\; \varepsilon\, \sup |f|\:,
\eeq
where we used the notation~$\grave{\bf{\Psi}}_0^{\omega n} =
\bra \grave{\bf{\Psi}}^{\omega n}, \Psi_0 \ket$.

In order to estimate the contribution by~$U_+$, we use the positivity of~${{\bf{U}}}^{\omega n}_+$,
\begin{eqnarray}
\lefteqn{
\sum_{n \geq n_0}
\left| \int_{-\infty}^\infty f(\omega) \, \langle \grave{\bf{\Psi}}_0^{\omega n}, {{\bf{U}}}^{\omega n}_+
\grave{\bf{\Psi}}_0^{\omega n} \rangle_{\C^2} \right|
\;\leq\; \sup|f| \sum_{n \geq n_0} \int_{-\infty}^\infty \langle \grave{\bf{\Psi}}_0^{\omega n}, {{\bf{U}}}^{\omega n}_+ \grave{\bf{\Psi}}_0^{\omega n} \rangle_{\C^2} } \nonumber \\
&=& \sup |f|
\left( \sum_{n \geq n_0} \int_{-\infty}^\infty \langle \grave{\bf{\Psi}}_0^{\omega n}, {{\bf{U}}}^{\omega n} \grave{\bf{\Psi}}_0^{\omega n} \rangle_{\C^2} -
\sum_{n \geq n_0} \int_{-2J}^{2J} \langle \grave{\bf{\Psi}}_0^{\omega n}, {{\bf{U}}}^{\omega n}_- \grave{\bf{\Psi}}_0^{\omega n} \rangle_{\C^2} \right) . \label{modes}
\end{eqnarray}
Using the asymptotic form near infinity of the energy density of the initial wave packet, an explicit computation
shows that
\begin{eqnarray*}
\lim_{L \rightarrow \infty} \bra \Psi_0, \Psi_0 \ket &=&
\lim_{L \rightarrow \infty} \int_{-\infty}^\infty \langle \grave{\bf{\Psi}}_0^{\omega \tilde{n}}, {{\bf{U}}}^{\omega \tilde{n}} \grave{\bf{\Psi}}_0^{\omega \tilde{n}} \rangle_{\C^2} \\
\lim_{L \rightarrow \infty} \grave{\bf{\Psi}}_0^{\omega n} &=& 0 \qquad
{\mbox{for every~$n \neq \tilde{n}$}}.
\end{eqnarray*}
It follows that
\begin{eqnarray*}
\lefteqn{ \lim_{L \rightarrow \infty} \sum_{n \geq n_0} \int_{-\infty}^\infty \langle \grave{\bf{\Psi}}_0^{\omega n}, {{\bf{U}}}^{\omega n} \grave{\bf{\Psi}}_0^{\omega n} \rangle_{\C^2}
\;=\; \lim_{L \rightarrow \infty} \sum_{n \neq \tilde{n}} \int_{-\infty}^\infty \langle \grave{\bf{\Psi}}_0^{\omega n}, {{\bf{U}}}^{\omega n} \grave{\bf{\Psi}}_0^{\omega n} \rangle_{\C^2} } \\
&=& \lim_{L \rightarrow \infty} \left( \bra \Psi_0, \Psi_0 \ket -
\int_{-\infty}^\infty \langle \grave{\bf{\Psi}}_0^{\omega \tilde{n}}, {{\bf{U}}}^{\omega \tilde{n}} \grave{\bf{\Psi}}_0^{\omega \tilde{n}} \rangle_{\C^2} \right) \;=\; 0 \:.\spc\spc
\end{eqnarray*}
Using this estimate in~(\ref{modes}), we conclude that
\begin{eqnarray*}
\lefteqn{ \limsup_{L \rightarrow \infty}
\sum_{n \geq n_0}
\left| \int_{-\infty}^\infty f(\omega) \, \langle \grave{\bf{\Psi}}_0^{\omega n}, {{\bf{U}}}^{\omega n}_+
\grave{\bf{\Psi}}_0^{\omega n} \rangle_{\C^2} \right| } \\
&\leq& \sup|f| \,\limsup_{L \rightarrow \infty}
\sum_{n \geq n_0} \left| \int_{-2J}^{2J} \langle \grave{\bf{\Psi}}_0^{\omega n}, {{\bf{U}}}^{\omega n}_- \grave{\bf{\Psi}}_0^{\omega n} \rangle_{\C^2} \right|
\;\leq\;  \varepsilon \sup|f| \:,
\end{eqnarray*}
where in the last step we have used~(\ref{Umies}).
\QED

\noindent
{\em{Proof of Theorem~\ref{thm2}. }}
We again consider the energy splitting for the individual angular momentum modes,
\[ \Psi_n^\low(t) \;=\; (e^{-i t H}\, \chi_\low(H))_n \Psi_0, \qquad
\Psi_n^\high(t) \;=\; (e^{-i t H}\, \chi_\high(H))_n \Psi_0 \:. \]
Using the identity~$1 = (\chi_\low + \chi_\high)^2$,
we obtain in analogy to~(\ref{psihighes}) that
\[ \sum_{n \geq n_0} \bra \Psi_0, (\chi_\high^2)(H)_n \Psi_0 \ket
\;=\;
\sum_{n \geq n_0} \bra \Psi_0, (1 - \chi_\low (\chi_\low + 2 \chi_\high))(H)_n \Psi_0 \ket \:. \] 
Applying again~\cite[Theorem~1.1]{FKSY2} gives the estimate
\[ \left\| \sum_{n \geq n_0} \Phi_n^\high \right\|_{L^2}^2 \;\leq\;
c \sum_{n \geq n_0} \bra \Psi_0, (1 - \chi_\low (\chi_\low + 2 \chi_\high))(H)_n \Psi_0 \ket \:, \]
where~$c$ is independent of~$\Psi_0$. Applying Lemma~\ref{lemma83}, by choosing~$n_0$
sufficiently large, the right side can be made arbitrarily small uniformly in~$L$.
Combining this with Lemma~\ref{lemma82}, we conclude that for any~$\varepsilon$
there is~$n_0$ such that
\[ \left\| \sum_{n \geq n_0} \Phi_n \right\|_{L^2((-\infty, 2 u_0) \times S^2)}
\;\leq\; \varepsilon \]
uniformly for large~$L$. Applying this estimate to the time derivatives
and using the equation together with the fact that the spatial part of the wave
operator is uniformly elliptic, we get similar estimates for the integral
of the energy density near the event horizon.
From the local decay~\cite{FKSY2}, we know that (for any fixed~$L$) the energy in any compact set
tends to zero as~$t \rightarrow \infty$.
Furthermore, we know from Lemma~\ref{lemma83} that the total energy of the
high angular momentum modes is small uniformly in~$L$.
Putting these facts together, the theorem follows.
\QED

\noindent
{\em{Acknowledgments:}} We are grateful to the Alexander-von-Humboldt Foundation 
and the Vielberth Foundation for their generous support.
We thank Niky Kamran and Shing-Tung Yau for their interest and encouragement.

\begin{tabular}{lcl}
\\
Felix Finster & $\;\;\;\;$ & Joel Smoller \\
NWF I -- Mathematik && Mathematics Department \\
Universit{{\"a}}t Regensburg && The University of Michigan \\
93040 Regensburg, Germany && Ann Arbor, MI 48109, USA \\
{\tt{Felix.Finster@mathematik.uni-r.de}} && {\tt{smoller@umich.edu}}
\end{tabular}

\end{document}